\documentclass[a4paper, 11pt, onsided, twocolumn]{article}
\usepackage[utf8]{inputenc}

\usepackage{graphicx}
\usepackage{authblk}
\usepackage[a4paper,width=180mm,top=20mm,bottom=20mm,bindingoffset=0mm]{geometry}
\usepackage{siunitx}
\DeclareSIUnit\angstrom{\text {Å}}
\usepackage[version=3]{mhchem}
\usepackage{amssymb}
\usepackage{pgffor}
\usepackage[super,sort&compress,comma]{natbib} 
\usepackage[hidelinks]{hyperref}
\usepackage[section]{placeins}
\usepackage{pdflscape}
\usepackage[font=footnotesize,labelfont=bf]{caption} 
\usepackage{enumerate}

\usepackage{xspace}
\usepackage[T1]{fontenc}

\usepackage{xcolor}

\title{\textit{In situ vs ex situ}: Comparing the structure of PNIPAM microgels at the air/water and air/solid interfaces}

\author[a]{H. Robertson}%
\author[a]{J. Zimmer}%
\author[a]{A. Sifuentes Name}%
\author[a]{C. Lux}%
\author[a]{S. Stock}%
\author[a]{R. von Klitzing}%
\author[a,*]{O.~Soltwedel}

\affil[a]{%
 Institut f\"ur Physik  Kondensierter Materie, Technische Universit\"at Darmstadt, D-64289 Darmstadt, Germany
}%
\affil[*]{olaf.soltwedel@pkm-tu-darmstadt.de}

\usepackage{ xcolor }
\usepackage{subcaption}
\usepackage{siunitx}

\newcommand{\Angstrom}{\mbox{\normalfont\AA}}
\newcommand{\angstrom}{\mbox{\normalfont\aa}}
\usepackage{multicol}

\date{}

\begin{document}
\twocolumn[
\begin{@twocolumnfalse}
\maketitle
\begin{abstract}
For studying the structure of microgel particles at the air/water interface, specular and off-specular X-ray reflectivity (OSR/XRR) allows \textit{in situ} measurements without any labelling techniques. 
Herein we investigate the vertical and lateral structure of poly(\textit{N}-isopropylacrylamide) (PNIPAM) microgels (MGs) at the air/water interface and the effect of Langmuir-Blodgett (LB) transfer onto solid substrates. 
The initial \textit{ex situ} atomic force microscopy (AFM) scans of LB-transferred MGs at the air/solid interface reveal strong lateral 2D hexagonal ordering across a broad range of lateral surface pressures at the air/water interface before LB-transfer. 
Notably, for the first time, these results were confirmed by OSR, demonstrating the existence of the long-range hexagonal ordering at low and intermediate surface pressures. 
For \textit{in situ} conditions and upon uniaxial compression at the air/water interface, the MG lattice constant decreases non-monotonically. This indicates the formation of domains at low pressures that approach each other and only compress when the surface isotherm reaches a plateau. 
Comparing the results of \textit{in situ} and \textit{ex situ} measurements, our study demonstrates a clear transfer effect during the LB-deposition on the lateral ordering of the MGs: the distance between the particles decreased during LB-transfer, and at high pressures ($\Pi\,>\,17\,\mathrm{mNm^{-1}}$) a second distance occurs indicating small domains with hexagonal internal ordering. 
The novel surface characterisation approaches debuted here highlight the use of both XRR and OSR to probe the vertical and lateral structure of adsorbed MGs, offering \textit{in situ}, non-invasive insights without the need for doping or transfer-induced artefacts.
\end{abstract}
\end{@twocolumnfalse}
]

\section{Dedication}

Through this article, we would like to contribute to the themed issue on the Memorial volume for Prof.~Dr.~Stefan Egelhaaf (1963-2023), our highly esteemed colleague. He shaped the field of particle ordering in bulk solutions \textit{via} scattering experiments and is well-known and respected for his remarkable contributions to the deep understanding of the relationship between structure and rheology in colloid science. 

\section{Introduction}
Colloidal particles, such as soft polymer microgels (MGs) or hard, partially hydrophobised silica nanoparticles, can self-assemble to form monolayers at interfaces. 
Such monolayers are ubiquitous in nature and industry and hold great interest from a fundamental perspective.\cite{Scheffold2020} Notably, the chemical properties of these colloidal particles can be tailored to modify self-assembly, and in turn, modify interfacial properties such as surface tension and adhesion. \cite{Xu2022,Forg2022}
Using a Langmuir trough (LT), monolayers can be produced by the immobilisation of colloidal particles at the interface of two immiscible fluids; e.g., oil/water or air/water.
Lateral trough barriers constrain the particle concentration per unit area and thus also the surface pressure ($\Pi$), allowing for the \textit{in situ} investigation of the pseudo gas/liquid/solid phase transition. 

In addition, the structural arrangement of the colloidal species at the interface can be examined by monolayer transfer from the air/liquid interface to a solid substrate. 
This is known as Langmuir-Blodgett (LB) deposition. 
In this regard, the adsorption of polymer microgels to various interfaces has been widely studied as a model system.\cite{2024Kuk} 
Polymer microgels, such as those composed of thermoresponsive poly(\textit{N}-isopropylacrylamide) (PNIPAM), are three-dimensional polymeric networks and spontaneously adsorb at the oil/water and air/water interfaces.\cite{Zhang.1999}
In contrast to hard colloidal particles, soft MGs deform during adsorption at the liquid interface and may adopt their shape in confinement. 
Previous studies have investigated the ordering of MG particles as a function of increasing surface pressure at the air/water interface mostly \textit{via} LB-transfered MGs on a solid substrate and atomic force microscopy (AFM).
Interestingly, they revealed the occurrence of an isostructural solid/solid phase transition in which MG particles with a high cross-linker content (i.e., stiffer MGs) arrange into two hexagonal crystalline phases with two lattice constants at a \textit{quasi}-constant lateral pressure.\cite{Rey.2016,Picard.2017,Rey.2017}
Nonetheless, due to the monolayer transfer step and substrate drying, the LB-deposition is considered an \textit{ex situ} method which does not directly examine the MG ordering at the air/water interface. 

In a recent study by Kuk~et~al., the MG arrangement at the air/water interface was investigated \textit{in situ} by a LT with small-angle light scattering (SALS).\cite{2023Kuk} 
By measuring diffraction patterns with increasing surface pressure, the real space interparticle distance was calculated, revealing no occurrence of an isostructural solid/solid phase transition as determined by the \textit{ex situ} measurements. 
The authors concluded that an \textit{ex situ} investigation \textit{via} LB-deposition can drastically change the structural properties of the colloidal monolayers; attractive capillary forces arising during substrate transfer and drying could lead to the rearrangement of MG particles or the formation of clusters.\cite{Denkov.1992, Kralchevsky.2000, Vogel.2012}
On hydrophilic substrates (e.g., silicon wafers), as used in most transfer studies, the attraction between microgel particles and the substrate is weak relative to that on hydrophobic substrates.\cite{2024Kuk, HoppeAlvarez2021}
These hypotheses are supported by observations that the drying conditions affect the observed microstructure; ``slow’’ substrate drying leads to the discussed isostructural solid/solid phase transition, whereas no such transition was observed in ``fast’’ drying.\cite{2024Kuk}

Neutron reflectometry experiments have recently revealed the structure of MG particles at the air/water interface below and above the volume phase transition temperature,\cite{2022Bochenek} as well as upon lateral compression of MG particles with various cross-linker densities.\cite{2024Gerelli} 
Using air contrast matched water, the latter work deduced the polymer volume fraction along the surface normal. 
However, one limitation of this protocol was that due to the lack of contrast, the authors were unable to determine the corresponding water fraction, and with this, the liquid/surface topology. 
Kuk~et~al. studied large MG particles (diameter of \SI{1}{\micro\metre}) with a silicon core of (diameter of \SI{0.5}{\micro\metre}) to gain contrast against the bulk phase for light scattering. This drastically increases the rigidity of the MG-cores and enhances the resulting capillary forces that do not occur for soft particles.\cite{2023Kuk} 
However, here we aim to compare the structural properties of smaller, and therefore softer, MGs at the air/water interface (\textit{in situ}) and immobilised at the solid/air interface (\textit{ex situ}). In order to obtain lateral structural information in the $\mathrm{\mu m}$ range, without artificial contrast, at various interfaces, we introduce off-specular X-ray scattering (OSR). In addition, we also study the vertical structure of the MG monolayer employing specular X-ray reflectivity (XRR) to track deformation along the surface normal upon compression.\cite{Rey.2023, ROBERTSON2024}

Herein, for the first time, we employ both specular and off-specular X-ray reflectivity (XRR) to characterise the structural properties of PNIPAM MG particles at the air/water interface by tracking changes in electron density across the interface. 
Specular XRR is employed to determine the \textit{in situ} vertical MG volume fraction profile and is complemented by corresponding \textit{ex situ} AFM micrographs. Additionally, we debut the use of off-specular XRR to monitor the lateral structure of the MG-decorated air/water interface as a function particle density. Importantly, we demonstrate that a laboratory X-ray source is sufficient to investigate these inhomogeneous MG particle films, providing an explanation for the unexpected non-homogeneous behaviour.

\section{Experimental}
\subsection{Materials}
\textit{N}-isopropylacrylamide (NIPAM, $(\geq)\,99\%)$ and \textit{N,N}'-methylenbisacrylamide (BIS, $99\%$) were purchased from Sigma-Aldrich (Merck, Darmstadt, Germany) and used as received. 
2,2'-azobis(2-methyl-propanimidamide) dihydrochloride (AAPH, $98\%$) was purchased from Cayman Chemical Company (Cayman Chemical, USA). Purified water (specific resistivity $\SI{18.2}{\mega\Omega\cdot\centi\meter}$ at $\SI{25}{\celsius}$) was obtained by a Milli-Q purification system (Merck KGaA, Darmstadt, Germany). 
Ethanol (EtOH) was purchased from Carl Roth GmbH + Co. KG, Karlsruhe, Germany.    

\subsection {Microgel synthesis and bulk characterisation}
The PNIPAM microgels were synthesised by surfactant-free precipitation polymerisation:\cite{1986Pelton} \SI{19.6}{mmol} NIPAM and \SI{0.4}{mmol} BIS (corresponding to a crosslinking density of \SI{5}{mol\%}) were dissolved in \SI{120}{mL} of \ce{H2O} and transferred into a home-build, double-walled glass reactor.  
The mixture was degassed with \ce{N2} for 60~min at \SI{80}{\celsius}. Subsequently, 0.125~mmol of AAPH dissolved in 1~mL H$_{2}$O were injected into the reaction mixture under constant stirring at 1000~rpm for the initiation of the polymerisation reaction. Stirring was terminated after 90~min and the turbid solution cooled to room temperature. For removal of undesired synthesis residues, the microgel suspension was dialysed for 10~days against Milli-Q H$_{2}$O, with a daily water change. Following this, four sedimentation (\textit{via} centrifugation 10000$\cdot$g, 30~min)-redispersion cycles were carried out. The obtained microgel suspension was lyophilised and the dried microgels were stored at \SI{-20}{\celsius} before usage. 

The hydrodynamic radius of the microgel particles (0.006~wt\% in H$_{2}$O) was determined by dynamic light scattering (DLS) measurements using a multi-angle light scattering instrument (LS Instruments AG, Fribourg, Switzerland) at 22 and \SI{50}{\degree}.
The correlation functions were measured in water at 9 angles (\SIrange{40}{120}{\degree} in \SI{10}{\degree} steps), resulting in a hydrodynamic diameter of \SI{616}{\nano\metre} at \SI{22}{\degree} and \SI{269}{\nano\metre} at \SI{50}{\degree}. \cite{Stock2022}

\subsection{Compression isotherms and Langmuir-Blodgett deposition} 

The compression isotherms were recorded at \SI{20}{\celsius} using the RK1 standard Langmuir trough ($A_\mathrm{min} = 15\,\mathrm{cm}^{2}$, $A_\mathrm{max} = 200\,\mathrm{cm^{2}}$) ($\mathrm{Riegler\,\&\,Kirstein}$ GmbH, Potsdam, Germany). A Wilhelmy tensiometer equipped with blotting paper was used to determine the surface pressure. The MG sample had been prepared beforehand by mixing the respective MG dispersion (1~wt\% in H$_{2}$O) with EtOH (20 vol\%). Before each measurement, the trough was thoroughly cleaned with chloroform and water. Then the trough was filled with $\mathrm{130\,mL}$ of $\mathrm{H_2O}$, with expanded barriers. Using a Hamilton syringe an exact amount of MG sample was spread dropwise at the air/water interface. To allow the EtOH to evaporate and ensure an equilibrated MG film at the air/water interface, the film was left to equilibrate for $\mathrm{30\,min}$ prior to compression. The layer was compressed with a constant speed of $\mathrm{34\,mm^{2}s^{-1}}$ and the lateral pressure ($\Pi$) was recorded simultaneously. In the resulting ${\Pi-A}$ isotherms the area was normalised to the spreading mass.

The Langmuir-Blodgett deposition of the MG particles was performed on a silicon substrate (1~$\times$~3.5~cm$^2$) with a natural oxide layer.
Prior to deposition, the substrate was cleaned through immersion in EtOH and treated in an ultrasonic bath for 15 min and then dried with N$_{2}$. 
For the deposition, the wafer was fixed to a home-built dip device keeping them in the water phase inclined by \SI{25}{\degree} towards the air/liquid interface. The MG monolayer was then prepared in the same way as explained above. When the targeted surface pressure was reached, the system was left to equilibrate for another 30 min before the silicon wafer was lifted through the interface with a constant speed of $\mathrm{25\,mm^{2}s^{-1}}$. The trough feedback system allowed for continuous adjustment of the barrier positions, which ensured a constant lateral pressure. The slow movement of the slightly inclined wafers allowed them to dry during the deposition process; i.e., without further treatment. All samples were prepared at a constant temperature of \SI{20}{\celsius}. 

\subsection{\textit{In situ} X-ray reflectivity} 

Specular X-ray reflectivity (XRR) and off-specular X-ray reflectivity (OSR) experiments pertaining to \textit{in situ} samples were conducted on a D8-Advanced (Bruker, Karlsruhe, Germany) diffractometer. 
A solid state Cu-K$\alpha$ anode ($\lambda = 1.54\,\mathrm{\Angstrom}$) was employed as a source and operated at \SI{40}{KV}, \SI{40}{mA}. 
Beam monochromation and collimation were achieved \textit{via} a Goebel mirror and a pair of slits (each \SI{0.1}{mm} opening and separated by \SI{10}{cm}). 
An enclosed trough was placed at the sample position (Riegler~\&~Kirstein~GmbH, Potsdam, Germany), equipped with a single barrier and dimensions of $11.5\,\mathrm{cm}$ parallel and $32.5\,\mathrm{cm}$ perpendicular to the beam propagation, covering $A_\mathrm{min} = 80\,\mathrm{cm}^{2}, A_\mathrm{max} = 320\,\mathrm{cm}^{2}$, reaching a compression ratio of $1:4$. 
The lateral pressure was monitored during compression using a Wilhelmy plate at approximately \SI{25}{\celsius}. 
The entire sample stage was actively vibration isolated (TS-140, The Table Stable Ltd., Mettmenstetten, Swiss). 
The microgel monolayers for \textit{in situ} X-ray scattering were prepared by spreading ethanol-dissolved MG on water. Together with the MG mass concentration of the ethanol solution, the deposited volume and the trough area were recalculated as the area per mass. A constant compression speed of $\mathrm{34\,mm^{2}s^{-1}}$ was employed. 
Once the desired lateral pressure was reached, the (off-)specular reflectivity was recorded at constant lateral pressure.

For the specular reflectometry measurements, we employed a 0-dimensional detector (NaI) and additional collimation slits (first $0.2~\mathrm{mm}$, second $0.1~\mathrm{mm}$ openings, $10~\mathrm{cm}$ separated from each other) on the secondary arm. 
To compensate for the rapid decay in reflectivity the counting time was increased stepwise with $q_\mathrm{z}$, resulting in roughly $40\,\mathrm{min}$ data acquisition time per reflectivity profile.

For the off-specular measurements, the beam collimation on the secondary arm was removed and a 1-dimensional detector ($\mathrm{V\angstrom ntec}$-1, spatial resolution $50~\mathrm{\mu m}$, 1600 channels) was installed. 
A beam stop was also installed to block the intense specular reflected intensity at the used incidence angle ($\theta_i\,=\,0.15^\circ$); this resembles the critical edge of total external reflection for the air/water interface. 
In this configuration, the detector covers the range of \SIrange{0}{3.6}{\degree} achieving an angular resolution of $\Delta\theta_f = 0.0062^\circ$. 
Data acquisition time for off-specular (detector) scan was $5\,\mathrm{min}$.

\subsection{\textit{Ex situ} X-ray reflectivity}

All \textit{ex situ} X-ray reflectometry measurements, i.e., those pertaining to Langmuir-Blodgett deposited microgels, were performed on a Rigaku SmartLab with a rotating copper anode ($\lambda=$\SI{1.54}{\angstrom}) and a Hybrid Pixel 2D detector under ambient conditions.
Off-specular measurements were performed at an incident angle of \SI{0.2}{\degree} and measuring the final angle across \SIrange{0.045}{7.358}{\degree} in 1D-mode; this translates to a $q_\mathrm{x}$-range of \SIrange{0}{0.0318}{\per\angstrom} in the beam propagation direction.
Beam collimation was achieved with a \SI{0.1}{mm} slit prior to sample irradiation. 
For the XRR, two additional \SI{0.1}{mm} slits (first mechanical, then post-sample illumination) are employed between sample and detector. 
To prevent over-illumination in the y-direction, the beam width was limited to \SI{5}{mm}, which does not improve the obtained $q_\mathrm{y}$ resolution significantly compared to the \textit{in situ} measurements.

Off-specular measurements were initially conducted with the substrate oriented in the same direction withdrawn from the trough during the Langmuir-Blodgett deposition process ($\phi\,=\,\SI{0}{\degree}$).
The sample was then rotated in \SI{2}{\degree} increments from \SIrange{0}{360}{\degree}. 
We admit that, due to the hexagonal symmetry, completing the full \SI{360}{\degree} rotation was not necessary, and \SIrange{0}{60}{\degree} would have been sufficient, however, we find it instructive for us and the reader to have the full picture, because in reality the sample footprint and misalignment could break this symmetry.  
At this stage, we emphasise the essential alignment of the surface normal that has to be parallel to the $\phi$ rotation axis. 
Off-specular reflection was measured for \SI{10}{\sec} at each angular position.

\subsection{AFM imaging and image analysis}
The AFM scans were carried out in ambient conditions with an MFP-3D-AFM (Oxford Instruments, UK) placed in an acoustic isolation enclosure. 
For scanning, an AC160TS cantilever (Olympus, Japan) with a nominal spring constant ($k$) of \SI{26}{N\per\meter} and resonance frequency (f$_{res}$) of \SI{300}{kHz} in tapping mode was used. 
Post-image processing involved flattening the height images for the removal of small sample tilts with respect to the AFM tip.
For the determination of the nearest neighbour distances by image analysis of the AFM scans, a custom-written Python script was used. 
Here, the respective AFM image is converted into a binary image and the position and number of centroids (particles) is determined.
For determination of the nearest neighbours of each centroid, the binary images are subjected to Voronoi tessellation, where only centroids that result in closed Voronoi cells are taken into account, as introduced by Rey~et~al.\cite{Rey.2016}
Subsequently, the nearest neighbouring distances were calculated. 
For each scan, the nearest neighbour distances were plotted as histograms and modelled with Gaussian functions.
The respective values displayed in Figure~\ref{pic:Dcc}B are the mean values of the Gaussian fits of the nearest neighbour distance histograms obtained each from the three independent sample positions at a given surface pressure. These scans were repeated at three independent sample areas.

\section{Data treatment \& analysis}

\subsection{Specular X-ray reflectivity - XRR}

Data were post-processed to correct for footprint, counting time, and attenuation settings for the individual scattering angles and merged into a single reflectivity curve. The wave vector transfer was calculated \textit{via} $q_\mathrm{z}\,=\,\frac{2\pi}{\lambda}(\sin{\theta_\mathrm{i}}+\sin{\theta_\mathrm{f}})$. Errors were estimated from the number of counts assuming a Poisson distribution. To highlight the impact of the surface layer, we normalised the data to the Fresnel reflectivity $R_\mathrm{F}$ of the substrate to counter the rapidly decaying reflectivity curve. The Fresnel reflectivity of the substrate (here water) was calculated according to $R_\mathrm{F}\,=\,\lvert\frac{\theta_\mathrm{i}-\theta_\mathrm{t}}{\theta_\mathrm{i}+\theta_\mathrm{t}}\rvert^2$ with $\theta_\mathrm{i}$ being the incident angle, $\theta_\mathrm{t}\,=\,\arccos{\left(\frac{1}{n_\mathrm{H_2O}}\cos{(\alpha_\mathrm{i})}\right)}$ the angle of the transmitted beam and $n_\mathrm{H_2O}\,=\,1 -\delta$ is the index of refraction of water for X-rays of the used wavelength. Here the dispersion ($\delta$) is related to the classical electron radius ($r_\mathrm{e}$) and the electron density of water ($\rho_\mathrm{e}$) \textit{via} $\delta\,=\,\frac{\lambda^2}{2\pi}r_\mathrm{e}\rho_\mathrm{e, H_2O}$.

The Fresnel-normalised reflectivity data were modelled indirectly by refining a scattering length density (SLD) profile from which the reflectivity is calculated recursively. 
Here we follow the phenomenological model approach successfully employed to refine neutron reflectometry data using 4 slabs.\cite{2022Bochenek} 
For all measurements along the $\Pi-A$ isotherm, the SLD profiles consisted of 4 boxes (or slabs), enumerated with the subscript $j\,= \,1,...4$ .
From fronting to backing, two slabs $(j = 1, 2)$ parametrises the floating part of the MG and another pair of slabs $(j = 3, 4)$ describes the part of the highly swollen MG adjacent to the apparent water sub-phase.
However, we introduced constraints between box thickness $l_\mathrm{j}$ and layer roughness $\sigma_\mathrm{j}$; to be precise $l_\mathrm{j}\,=\,2\sigma_\mathrm{j}$, to overcome ambiguous parameterisation for either thin parts or broad scattering length density distributions. Thus, all roughness parameters are coupled to the adjacent box thickness. To imitate the smooth decay of the electron density towards air, we coupled the length of the air adjacent boxes to each other ($l_1\,=\,l_2$). In total, this resulted in 7 fitting parameters (3 for thickness, 4 for scattering length density), the results of which are presented in Table~S3.1 in the SI. The numerical analysis is described by \citeauthor{1996Asmussen}\cite{1996Asmussen} and is based on the Parratt-Algorithm.\cite{1954Parratt} We would like to emphasise that the visibility of oscillations in the specular reflectivity is strongly enhanced by normalisation to the modelled reflectivity of a bare air/water interface; this enables the discrimination of layers that exhibit rough interfacial roughness.

\subsection{Off-specular X-ray reflectivity - OSR}

The $2\theta_\mathrm{f}$ resolved intensity was transformed \textit{via} $q_\mathrm{x}\,=\,\frac{2\pi}{\lambda}(\cos{\theta_\mathrm{f}}\,-\,\cos{\theta_\mathrm{i}}$) and $q_\mathrm{z}$, taking into account the constant incident angle of $\theta_\mathrm{i}\,=\SI{0.15}{\degree}$ on water and $\theta_\mathrm{i}\,=\SI{0.2}{\degree}$ on silicon. Strictly mathematically speaking, in this convention, the reciprocal space is probed in the negative $q_\mathrm{x}$ direction, when $\theta_\mathrm{f}$ exceeds $\theta_\mathrm{i}$. Since the lattice is congruent when rotated every \SI{60}{\degree}, and thus also \SI{180}{\degree}, the $q_\mathrm{x}$ (and for completeness also the $q_\mathrm{y}$) axis can by mirrored without altering the result. Throughout this entire work we exploit this reflection symmetry, using only the absolute value of $q_\mathrm{x}$ and limit the data analysis for OSR on the scattered intensity above the specular condition (i.e., $\theta_\mathrm{f}~>~\theta_\mathrm{i}$).

With increasing $q_\mathrm{x}$, the off-specular reflected intensity decays roughly with $q_\mathrm{x}^{-2}$. This scaling is mathematically described by the static structure factor for a finite two-dimensional harmonic lattice. \cite{Dutta1981} 
In a more general framework, the differential cross section $\frac{d\sigma}{d\Omega}\,\sim\,\frac{k_\mathrm{B}T}{\gamma q_\mathrm{x}^2}\left(\frac{q_\mathrm{x}}{q_\mathrm{ max}}\right)^\eta$ shows the same result. 
Here $k_\mathrm{B}, T, \gamma$, and $q_\mathrm{max}$ are, respectively, the Boltzmann constant, temperature, surface tension and the Debye cutoff. 
The latter stems from the shortest wavelength of the capillary waves, with a realistic value of $q_\mathrm{max}\,=\,\SIrange{1.2}{2.5}{\Angstrom^{-1}}$ for a free air/water interface, that governs the diffuse scattering.\cite{Braslau1988} Since the dimensionless parameter $\eta\,=\,\frac{k_\mathrm{B}T}{2\pi\gamma} q_\mathrm{z}^2$ depends on $q_\mathrm{z}^2$, the decay of OSR follows not exactly a scaling law, yet is in good approximation small, for the here covered reciprocal space $q_\mathrm{z}\,=\,\SIrange{0}{0.2}{~\Angstrom^{-2}}$, and calculates to $\eta\,\leq\,0.2$. For simplicity, we will therefore re-normalise the data by the factor $q_\mathrm{x}^{-1.8}$. 

\begin{figure}[h]
\centering
  \includegraphics[width=\columnwidth]{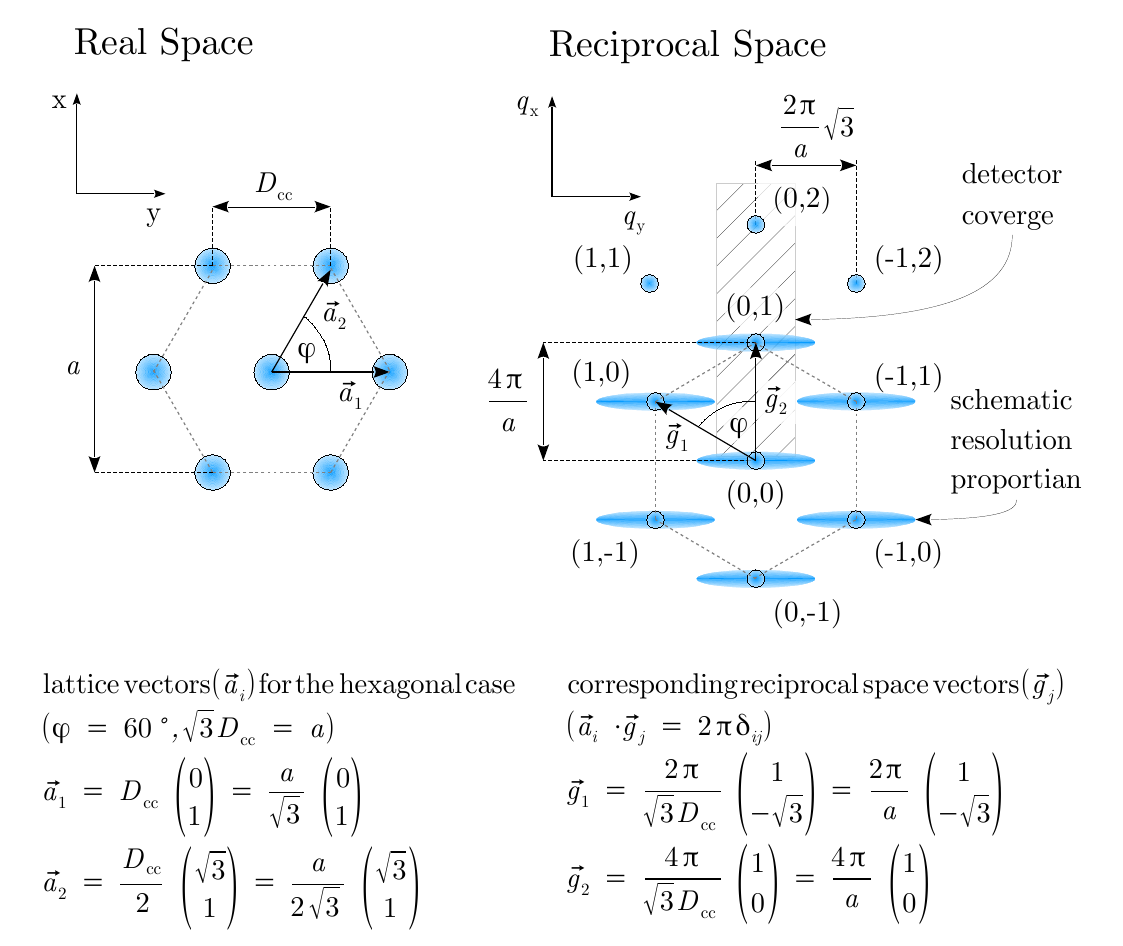}
\caption{Artistic representation for a 2D-hexagonal lattice in (left) real and (right) reciprocal space, together with the used convention for lattice vectors and indicated resolution in reciprocal space.}
\label{pic:Scheme_2D-Lattice}
\end{figure}

From literature and our AFM-investigation we assume a 2D-hexagonal ordering of the microgels at the air/water interface. The situation is schematically depicted in  Figure~\ref{pic:Scheme_2D-Lattice} (left) as a birds-eye view in real space. Here the centre to centre distance ($D_\mathrm{cc}$) between the MG particles is indicated. Its absolute value equals the real space vectors $\vec{a_\mathrm{1}}$ and $\vec{a_\mathrm{2}}$. The topology of the adsorbed MG is paraphrased in the literature as a ``fried egg'' shape, which describes an extended flat (2D) corona (egg white) around a hemispherical core (egg yolk).\cite{Mourran2016}

In addition to a constant factor, the re-normalised intensity is given in the Born approximation $I(q_\mathrm{x}) = |F(q_\mathrm{x},q_\mathrm{z}) \cdot S(q_\mathrm{x})|^2$, assuming that the structure factor only has an influence in the sample plane. Since the SLD difference is negligible between the water and the swollen microgel, we can only consider the part of the microgel above the apparent air/water interface to be described by a form factor ($F(q)$). The exact shape of $F(q)$ remains unresolved in this investigation, because we only use a single incidence angle ($\theta_\mathrm{i}$), which renders this endeavour impossible without additional constrains. Nevertheless, a detailed knowledge of the exact form factor is not a necessity to extract the structure factor. This becomes obvious when $F(q)$ does not alter the peak positions as function of $q_\mathrm{z}$. Within a proof of principle experiment we mapped the reciprocal space around the peak region (see Figure S2.4 in the SI) to confirm this point. The scattering pattern resembles crystal truncation or Bragg rods, also observed in surface X-ray diffraction from (\textit{semi})-infinite crystals or not tilted amphiphilic monolayers.\cite{Robinson1986, Kaganer1999} Here the scattering is characteristically diffuse in the direction perpendicular to the surface $(q_\mathrm{z})$, but sharp in parallel directions $q_\mathrm{x}$ and $q_\mathrm{y}$. At this point we would like to remind the reader of the strong asymmetric resolution of the employed scattering geometry in $q_\mathrm{x}$ and $q_\mathrm{y}$. While the scattering from ordered structures in the $x$-axis can be well resolved, the situations differs dramatically in the $y$-direction. We will return to this point after discussing the employed model for the structure factor.


In the first attempt, we employed a structure factor of 3 ($\Pi\,\leq\,15\,\mathrm{mNm^{-1}}$) or 4 ($\Pi\,>\,15\,\mathrm{mNm^{-1}}$) Gaussian functions. 
Their centre positions fulfil the 2D-hexagonal lattice spacing ($|\vec{q_\mathrm{h,k}}|\,=\,\frac{4\pi}{\sqrt{3}d}\cdot\sqrt{h^2+k^2+hk}$), having the position ratios $1, \sqrt{3},$ and $2$ for the first 3 Gaussian functions whilst permitting a $2\%$ fitting variation (for further details on the initial fitting approach the reader is referred to Figure~S2.1 and Table~2.1 in the SI). 
However, inspired by the (\textit{ex situ}) OSR measurements of the LB-transferred MG, we chose a different approach to treat the data.
The reason is that each hexagonally oriented MG domain gives origin to a Debye-Scherer ring (Figure~\ref{pic:DIP_AFM_OSR}A). 
Due to the poor $q_\mathrm{y}$ resolution, the umbrella effect is present,\cite{vanLaar1984} which is well known for the Kratky camera or, more generally, when slit collimation is used in Small Angle X-ray Scattering.\cite{Strobl1970} This resolution effect alters the measuring signal by (i) making the Bragg-Peak asymmetrical and (ii) shifts its position to smaller $q$-values.
More importantly, the periodicity of the hexagonal ordering generates pseudo peaks at $q$ positions smaller than the $1^\mathrm{st}$ order peak. 
To be precise, at $q^*\,=\,q_{(01)}\cdot\cos(\phi)$ with $\phi\,=\,30$~and~$60^{\circ}$, having the ratio $\sqrt{3/4}$ and $\sqrt{1/4}$.
The absence of the (1,1), (-1,2), ... peaks (see Figure~\ref{pic:Scheme_2D-Lattice}) that fulfil $|q| = \sqrt{3}|q_{(01)}|$ indicates that these lattice planes are distorted; this is reasonable as neighbouring MG particles are in contact with each other, reducing their distinctness.
For these reasons, it is sufficient to model the \textit{in situ} OSR with 6 (two sets of three) Gaussian functions, all linked to the lowest indexed peak with ratios given above; i.e., only one independent fitting parameter for the centre position of each of the 6 Gaussian functions.
A seventh unlinked Gaussian was employed as a heuristic parametrisation of the form factor.
The black Gaussian at the bottom of Figure~\ref{pic:OSR}A refers to the lowest indexed peak, the umbrella effect is addressed by the red Gaussians whose positions are fixed by the given ratios and the form factor is represented by the blue curve.
These results are presented in Figures~S2.2 and S2.3 in the SI.

\section{Results \& Discussion}
\subsection{Langmuir-Blodgett deposited microgels at the air/solid interface (\textit{ex situ})}

\begin{figure}
  \includegraphics[width=1\linewidth]{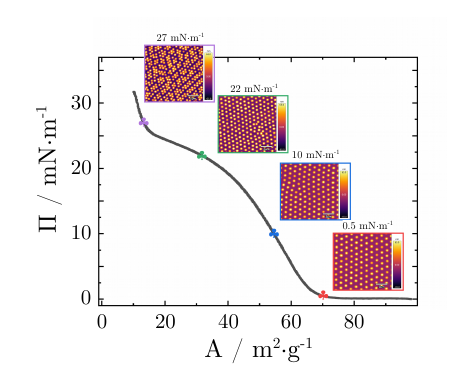}
\caption{Langmuir compression isotherm of the investigated microgel monolayers with representative AFM topologies of LB-transferred samples at the surface pressures $\Pi$ = 0.5, 10, 22, and 27 $\mathrm{mNm^{-1}}$. The AFM images are also shown in Figure~\ref{pic:DIP_AFM_OSR}C.}
\label{Dip positions}
\end{figure}

The \textit{ex situ} investigation of the structure of microgel particles at the air/water interface is conducted indirectly \textit{via} Langmuir-Blodgett deposition; i.e., microgel monolayer transfer from the liquid interface onto a solid substrate at defined surface pressures ($\Pi$).
These microgel films are then subsequently investigated with AFM.
The respective $\Pi$ values employed for deposition are $\Pi$ = 0.5, 10, 22, and 27~$\mathrm{mNm^{-1}}$, and their positions on the Langmuir isotherm are displayed in Figure~\ref{Dip positions}, whereas the OSR results and AFM micrographs are presented in Figure~\ref{pic:DIP_AFM_OSR}. 
The isotherm presents the characteristic shape already well described in literature.\cite{Pinaud2014, Picard.2017} In short, at $\Pi$=0~$\mathrm{mNm^{-1}}$, the MGs behave like a gas of non-interacting particles or clusters.
With decreasing area, a sharp increase in $\Pi$ is observed which corresponds to microgels in ``shell-shell'' contact and can be understood similarly to an expanded liquid phase.  
Upon further decreasing the area, $\Pi$ is seen to increase less drastically and even tends towards a plateau region, before another steep increase is observed as a result of microgel ``core-core'' interactions. 

\subsubsection{AFM (\textit{ex situ})}

For each surface pressure marked on the Langmuir isotherm in Figure~\ref{Dip positions}, an AFM scan of the corresponding LB-deposited sample is shown in Figure~\ref{pic:DIP_AFM_OSR}C. 
Here, image analysis of the AFM scans enabled the extraction of the distance between MG particles and their nearest neighbouring particles (D$_\mathrm{cc}$). 
These results are plotted as histograms in Figure~\ref{pic:DIP_AFM_OSR}D, where the interparticle distances were extracted by modelling the histograms with Gaussian functions.


As demonstrated in the AFM images at low surface pressures, i.e., $\Pi\,\le\,10\,\mathrm{mNm^{-1}}$ (Figure~\ref{pic:DIP_AFM_OSR}C), the MG particles arrange hexagonally over the scanned area. 
For these low surface pressures, the histograms were modelled with a single Gaussian, indicating that the MG arrangement can be characterised by one interparticle distance ($D_\mathrm{cc,\,AFM}$) which decreases from $0.98\,\pm\,0.06\,\mathrm{\mu m}$ ($\Pi\,=\, 0.5\,\mathrm{mNm^{-1}}$) to $0.89\,\pm\,0.08\,\mathrm{\mu m}$ ($\Pi\,=\,10\,\mathrm{mNm^{-1}}$). 
Increasing surface pressure to $\Pi{= 22}\mathrm{mNm^{-1}}$, the MGs are further compressed and the emergence of a second interparticle distance denoted as $D_\mathrm{agg}$, arises. 
This is visible in both the AFM image and the respective histogram, with $D_\mathrm{agg}\,=\,0.73\,\mathrm{\pm\,0.09 \mu m}$ and $D_\mathrm{cc,\,AFM}\,=\,0.62\,\pm\,0.08\,\mathrm{\mu m}$.  At $\Pi\,=27\,\mathrm{mNm^{-1}}$ the formation of MG ``aggregates'' is clearly observed in the AFM image, resulting in $D_\mathrm{agg}\,=\,0.70\,\pm\,0.13\,\mathrm{\mu m}$  and $D_\mathrm{cc,\,}\,=\,0.45\,\pm\, 0.09\,\mathrm{\mu m}$. 
Nevertheless, the ratio between the number of aggregates and the number of MG particles increases upon compression (see Figure \ref{pic:DIP_AFM_OSR}C). Langmuir-Blodgett deposition was also carried out at $\Pi\,=\,24,\,25,\, 26$, and 30~$\mathrm{mNm^{-1}}$ for the determination of the respective nearest neighbour distances. These results will be later discussed in \S5.3 (Figure~\ref{pic:Dcc}) and the respective AFM scans and corresponding histograms are presented in Figure~S1.1 in the SI. 
Additionally, Figure~S1.2 in the SI presents a stitch of four individual AFM scans, totally an area of \SI{100}{\micro\meter}~$\times$~\SI{100}{\micro\meter} which presents the distribution of MG particles within domains.

\subsubsection{OSR (\textit {ex situ})}

\begin{figure*}[t!]
\centering
\includegraphics[width=0.96\linewidth]{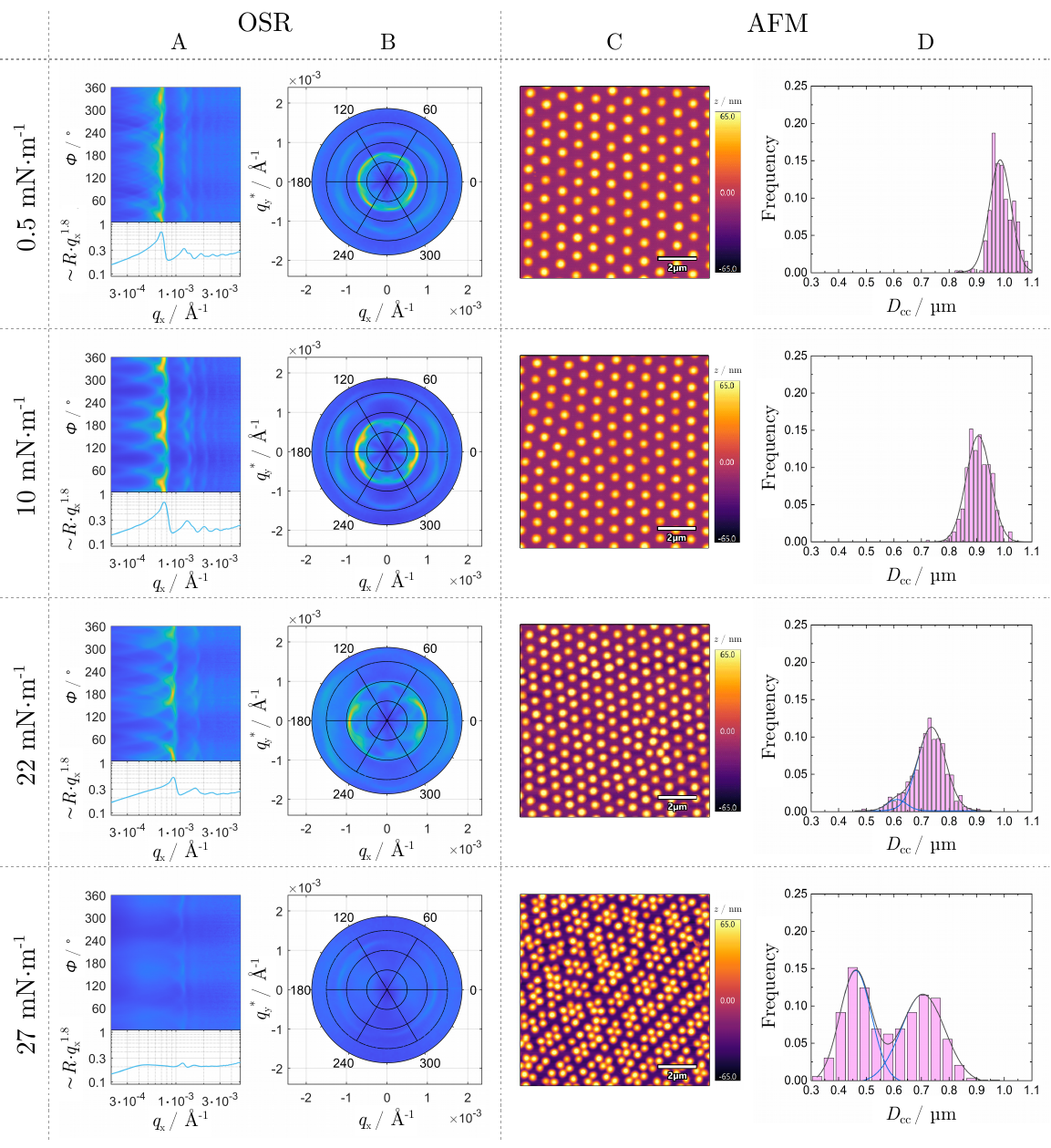}  
\caption{\textit{Ex situ} off-specular reflectometry (OSR) (columns A and B) and atomic force microscopy (AFM) (column C and D) data of PNIPAM microgels after Langmuir-Blodgett deposition onto a silicon substrate at surface pressures of $\Pi\,=\,0.5,\,10,\,22$ and $\SI{27}{mNm^{-1}}$ (see Figure~\ref{Dip positions}, from bottom to top). In column A, the orientation ($\phi$) resolved OSR (upper) and ($\phi$) integrated (lower) spectra are displayed in Cartesian coordinates. Corresponding polar coordinate representation with the same colour coding are presented in column B. Column C shows the AFM scans for the respective pressures. All topographies have the same colour-map and scan size. Corresponding histograms of the nearest neighbour distances $D_\mathrm{cc}$ shows column D, from which the mean interparticle distances are extracted by fitting the histograms using 1 (for $\Pi\,\leq\,10\,\mathrm{mNm^{-1}}$) or 2 ($\Pi\,>\,10\,\mathrm{mNm^{-1}}$) Gaussians.}
\label{pic:DIP_AFM_OSR}
\end{figure*}

The AFM micrographs clearly show a local hexagonal ordering of the adsorbed MG on the \si{\mu}m-length scale. 
However, to prove long-range ordering (across the whole $\mathrm{cm^2}$ sample) with AFM is not practical. 
To gain insight into the orientation distribution of the adsorbed MG particles over the entire sample, OSR measurements were performed as a function of the rotation around the sample z-axis ($\phi$). 
In addition to probing a larger statistical area, we can also compare results across both techniques. 
Due to the high in-plane sensitivity of the OSR technique (i.e., in the direction of the beam $q_\mathrm{x}$), the sample needs to be rotated around its surface normal by $\phi$ for a complete characterisation.   
The results are summarised in columns A and B of Figure~\ref{pic:DIP_AFM_OSR}, where the maximum intensity can be seen to shift towards larger $q_\mathrm{x}$ with increasing $\Pi$, indicating a decrease of the inter-particle distance $D_\mathrm{cc}$. 
Also, higher-order peaks are well detected, confirming large areas of coherent scattering. 

For $\Pi\,=\,27\,\mathrm{mNm^{-1}}$, the peak height is drastically reduced and no higher order peaks are visible.
This indicates a breaking of the long-range hexagonal ordering, aligning with the AFM results. 
For $\Pi\,<\,27\,\mathrm{mNm^{-1}}$, we observe a pronounced $\phi$ dependency of the OSR signals, whereby Debye-Scherrer ring-like structures are visible. 
Due to the hexagonal pattern of the MG particles, a minimum of 6 ring segments (i.e., every \SI{60}{\degree}) is present and is best distinguishable at $\Pi\,=\,10\,\mathrm{mNm^{-1}}$. 
Above and below $\Pi$~=~\SI{10}{mNm^{-1}}, the symmetry is less pronounced as additional orientations are present, which is a direct consequence of the available area for the MG particles. 
As the available area is larger than the MG particles can occupy, the domains can orientate around each other freely; for the reversed scenario, some neighbouring MG particles reduce their distance tremendously (Figure~\ref{pic:DIP_AFM_OSR}, column C at $\Pi\,=\,22\,\mathrm{mNm^{-1}}$), breaking away from the hexagonal symmetry that finally broadens the scattering signal.

It is worth noting that the superposition of the observed ring segments leads to pseudo-peaks at $q_\mathrm{x}$ positions below the lowest indexed peak. 
Due to the symmetry, the rings appear at multiples of $\cos(30)$ and $\cos(60)$ with respect to the lowest Bragg peak. These rings also contribute to scattering intensities below the Bragg peaks, making them asymmetric, as shown in the $\phi$-averaged OSR (lower panels of Figure~\ref{pic:DIP_AFM_OSR}, column A). The hexagonal symmetry (and its distortion) is best seen in the polar-coordinate figures (Figure~\ref{pic:DIP_AFM_OSR}, column B). 
From the position of the intensity maximum of the OSR signal ($\leq\,q_{01}$) one estimates the centre-to-centre distance ($D_\mathrm{cc}$) between the MGs by $D_\mathrm{cc}=\frac{4\pi}{\sqrt{3}q_{01}}$. 
This value is affected by the umbrella effect, which can increase the $D_\mathrm{cc}$ by several percent relative to the ``true'' value.
For $\Pi\,=\,10\,\mathrm{mNm^{-1}}$, the $q_\mathrm{01}$ peak is most intense at $\phi\,=\,0^\circ$ and $180\,^\circ$, indicating that the ordering along the $x$-axis (that is, (anti) parallel to the movement during deposition) is better ordered than the other direction. The preferred orientation of the immobilised MG particles is most probably introduced by the direction of the LB-transfer, because this is more invasive than the movement of the barriers and subsequent time for adsorption.
The same observation is made at $\Pi\,=\,22\,\mathrm{mNm^{-1}}$; here also only 6 (albeit blurred) ring segments are present. 
At the lowest pressure, numerous less intense circle segments are visible.
These subdued circle segments hint towards an incoherent superposition of various MG domains of different orientations; however, within each domain, MG particles are hexagonally ordered.

\subsection{Microgels at the air/water interface (\textit{in situ})}

\subsubsection{OSR (\textit {in situ})}

\begin{figure}[t!]
\centering
\includegraphics[width=1\linewidth]{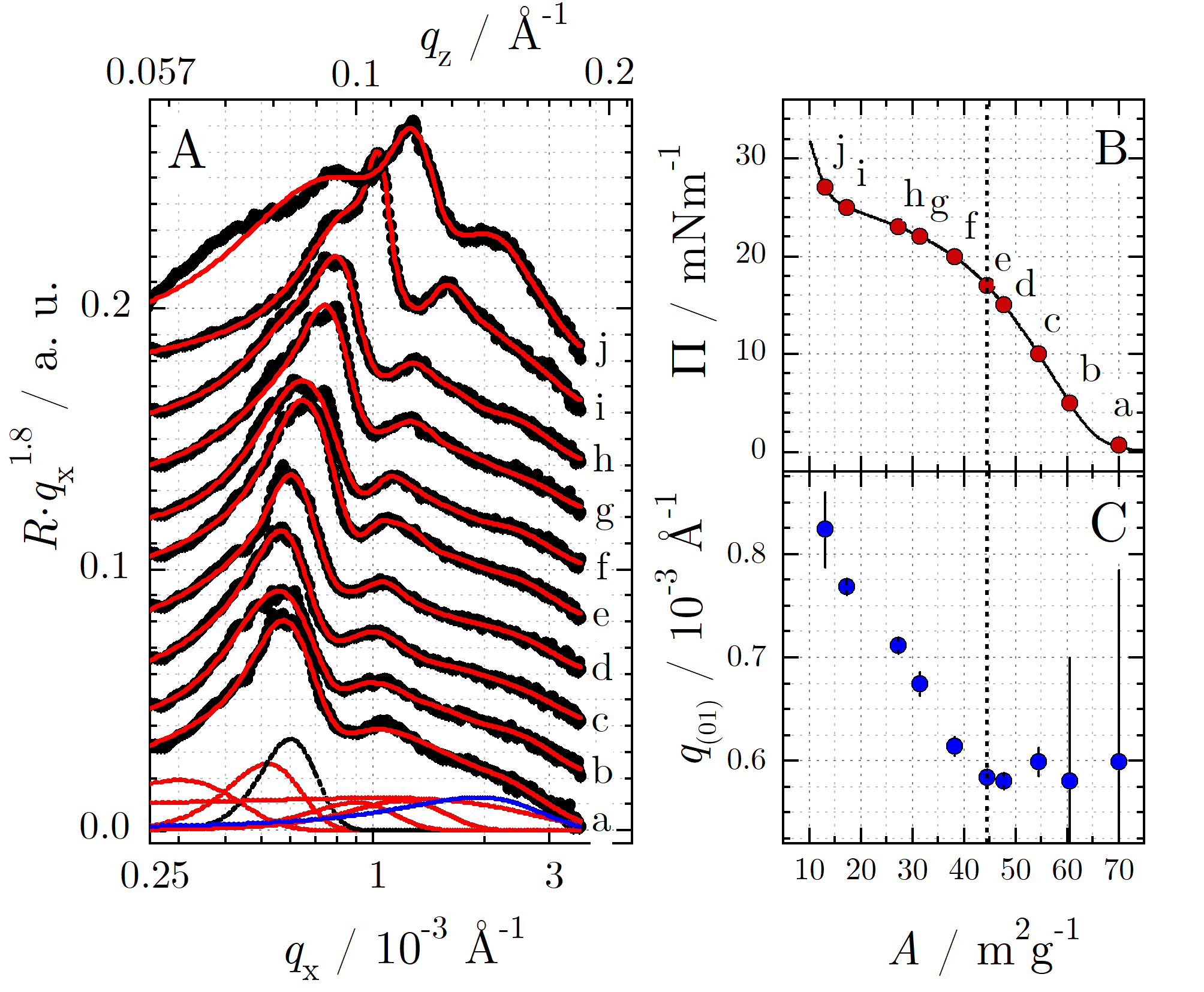}  
\caption{A: $q_\mathrm{x}^{1.8}$ weighted \textit{in situ} off-specular X-ray scattering of MG particles at the air-water interfaces at the pressures highlighted on the isotherm in panel B, starting from the bottom with the lowest lateral pressure. Black symbols are the measured values, and red solid lines are optimised models as described in the text. For improved visibility, the curves are shifted by $0.05$ against each other. B: $\Pi-A$ isotherm of the PNIPAM MG particles at the air/water interface (black solid line). Red symbols indicate the pressures at which the OSR were recorded.  C: Centre positions of the fitted Gaussian representing the (01) peak.}
\label{pic:OSR}
\end{figure}

Indeed a remarkable characteristic OSR is observable for all investigated pressures, as shown in Figure~\ref{pic:OSR}A. Here the data (black dots) and corresponding optimised models (red solid lines) are $q_\mathrm{x}^{1.8}$-weighted to compensate for the decay of the diffuse scattering. As an example, we have present the deconvoluted model for the spectra of the lowest pressure ($\Pi\,=\,0.5\,\mathrm{mNm^{-1}}$) in black and red, showing all Gaussians profiles that contribute to the structure factor. The black Gaussian profile corresponds to the (01) peak (and all others that have the same $|q|$ peak, see Figure~\ref{pic:Scheme_2D-Lattice}). The two red Gaussians at lowest $q_\mathrm{x}$ characterise the pseudo peaks, stemming from the umbrella effect of the first order Bragg-Peak. The remaining 3 Gaussian functions describe the second-order Bragg-Peak and corresponding pseudo peaks. A $7^\mathrm{th}$ Gaussian profile parameterises the background and the form factor, which is attributed to higher orders. The form factor contribution has only a marginal influence in our analysis and mainly contributes by a smooth decay with increasing $q$. The already explained model describes the data in the resolved $q_\mathrm{x}$-range astonishing well, with visible deviations starting at the highest lateral pressure. For the low-pressure region ($\Pi\,\le\,17\,\,\mathrm{mNm^{-1}}$), the peak centres remain almost invariant, while the pressure increases and the available area for the MG particles shrinks. Upon compression above $\Pi\,>\,17\mathrm{mNm^{-1}}$, the peak centre positions shift steadily to larger values, indicating a uniform shrinking of the lattice constant. At $\Pi\,=\,27\mathrm{mNm^{-1}}$, the centre position jumps remarkably, yet below this value, the model describes the OSR spectra at all pressures investigated. This suggests the presence of hexagonal structure with varying lattice constants (see panels B and C in Figure~\ref{pic:OSR}).

\subsubsection{XRR (\textit {in situ})}

\begin{figure}[t!]
\centering
\begin{subfigure}[t]{.5\columnwidth}
  \centering
  \includegraphics[width=1\linewidth]{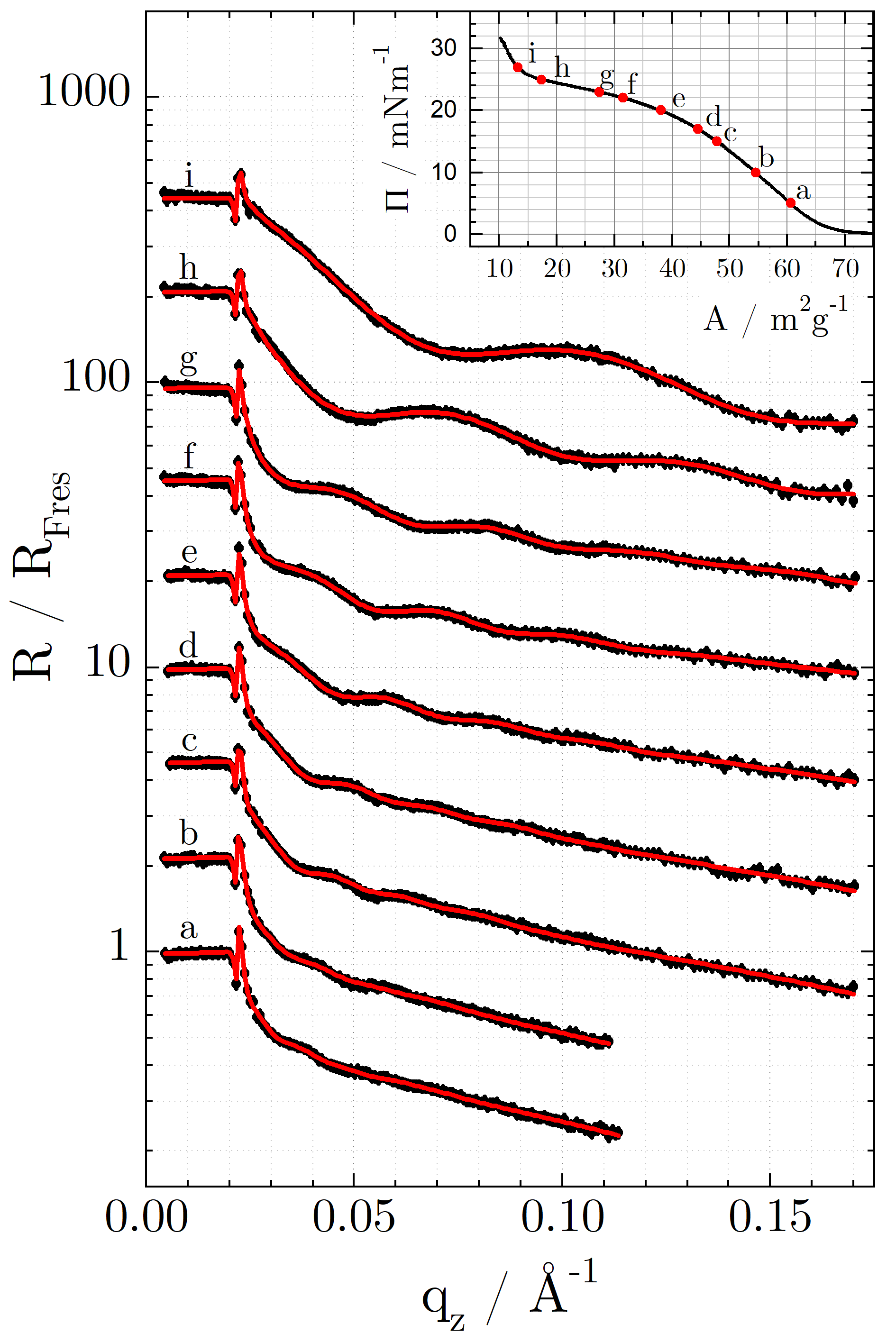}
  \end{subfigure}%
\begin{subfigure}[t]{.5\columnwidth}
  \centering
  \includegraphics[width=1\linewidth]{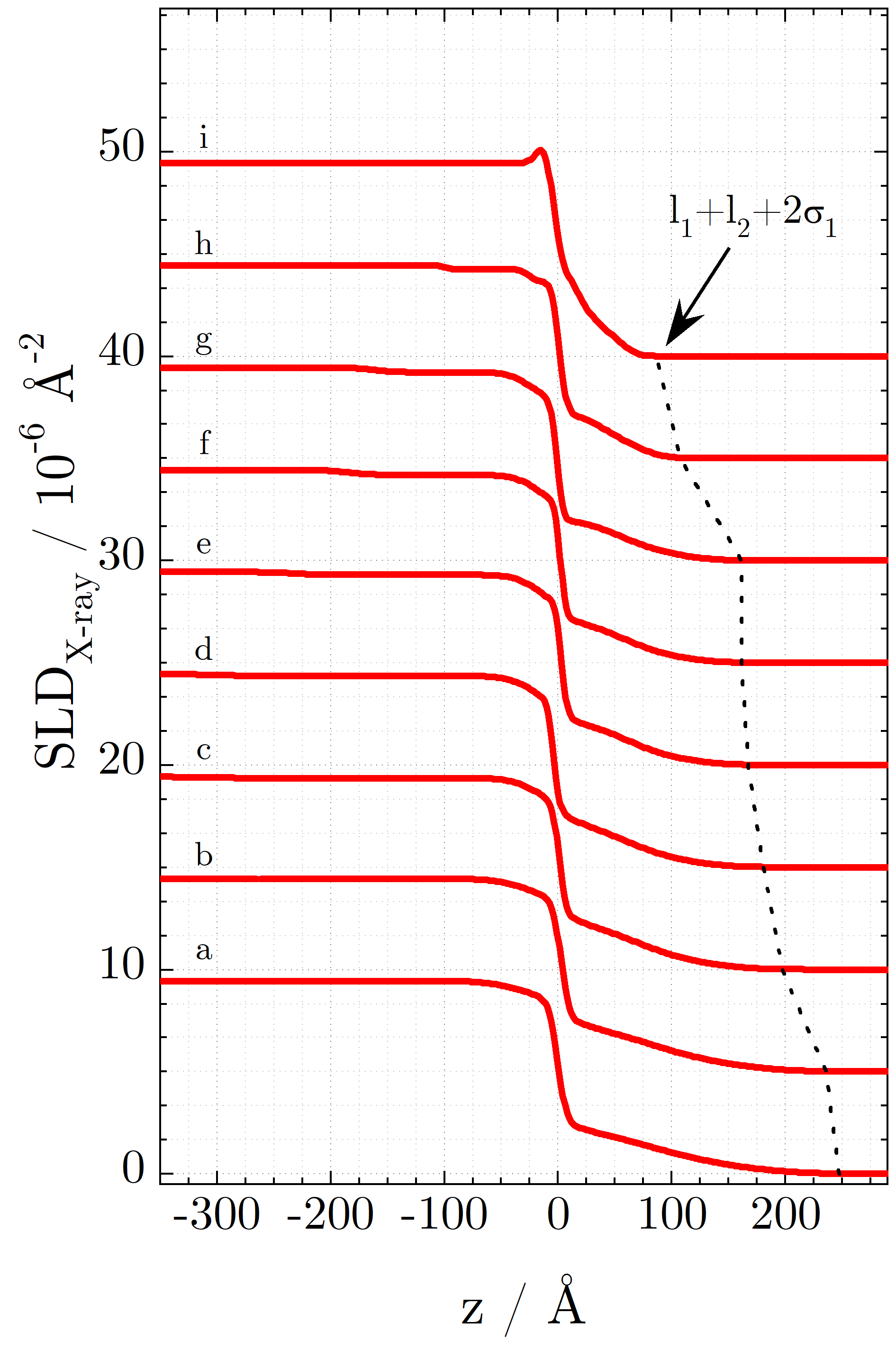}
  \end{subfigure}
\caption{(A) Fresnel normalised specular X-ray reflectometry of MG particles at the air/water interface and corresponding (B) SLD profiles at the pressures highlighted on the isotherm in the inset in A, starting from the bottom with the lowest lateral pressure. Black symbols are the measured values and red solid lines are the optimised models corresponding to the SLD profiles shown in (B). For improved readability, reflection profiles in (A) are offset by a factor of $\sqrt[3]{10}$ and SLD profiles in (B) are offset by $5 \cdot 10^{-6}~\Angstrom^{-2}$. The inset displays the sum of the length of the two slabs ($l_1$, $l_2$), including the adjacent transition region ($2\sigma_1$) to parameterise the layer present above the apparent air/water interface, as function of $\Pi$.}
\label{pic:XRR}
\end{figure}

Figure~\ref{pic:XRR} presents the \textit{in situ} XRR measurements of MG particles at the air/water interface together with the refined models (Figure~\ref{pic:XRR}A) and corresponding SLD profiles (Figure~\ref{pic:XRR}B) along the $\Pi$-A-isotherm. 
As the {Fresnel}-reflectivity ($R_\mathrm{Fres}$) is not convoluted with the instrumental resolution, it is infinitely sharp at the critical edge, leading to an artificial oscillation around this point. 
Nevertheless, the critical edge of total external reflection is constant throughout all measurements. All XRR data show a damped harmonic oscillatory decay with increasing $q_\mathrm{z}$. Upon lateral compression, the oscillations become more pronounced, with increasing amplitude, reduced periodicity, and less decay. The effects are most obvious for $\Pi\,\geq\,\,\SI{17}{mNm^{-1}}$ or (in analogy to lipid monolayers) when the liquid expanded phase is about to transition into the liquid condensed phase. These findings here are threefold: 
\begin{enumerate}[i)]
    \item the presence of a layer that thins upon compression; 
    \item the SLD value of the layers falls between the SLD of air and water; and 
    \item the layer SLD increases throughout all compression states. 
\end{enumerate}
This is confirmed by the modelled SLD profiles in Figure~\ref{pic:XRR}B. 
To further elucidate the origin behind the foremost finding, neutron scattering, which yields a higher contrast between the microgel particles and deuterated  solvents, is required.

Above the apparent air/water interface ($z\,=\,0\,\mathrm{\Angstrom}$), the MG film is parameterised by two boxes that describe the thicknesses ($l_1$, $l_2$), scattering length densities (SLD$_1$, SLD$_2$) and roughnesses ($\sigma_1$ - towards the air; $\sigma_2$ - between layers 1 and 2). 
The total thickness of this layer is given by $l_1\,+\,l_2\,+\,2\sigma_1$ or when taking the fit-constrains into account, simply $3l_1$. The change in layer thickness upon changes in surface pressure is indicated by the dashed line in the SLD profiles (Figure~\ref{pic:XRR}B) and plotted in the inset. 

\begin{figure}
\centering
  \includegraphics[width=\columnwidth]{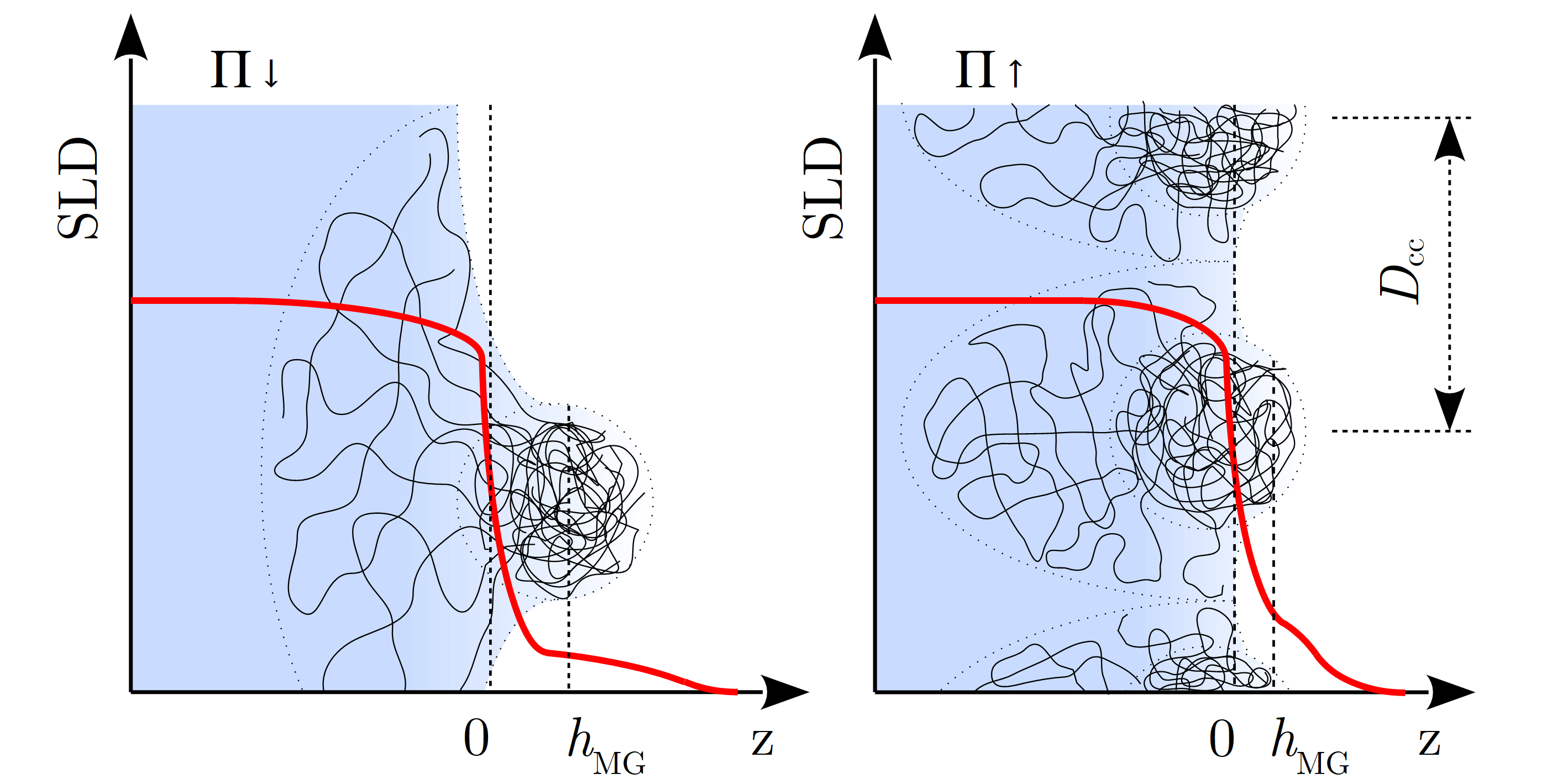}
\caption{Schematic representation of the structure of MG particles at the air/water interface at low (left) and high (right) lateral pressures. 
The resultant SLD profiles are superimposed in red along the surface normal.
The distance between microgel particles ($D_\mathrm{cc}$) and the height of the microgel above the air-water interface ($h_\mathrm{MG}$) are also denoted.
Note that in reality the ratio between $D_\mathrm{CC}$ and $z$ differs by 2-3 orders of magnitude, which is neglected in this artistically embellished illustration.}
\label{pic:Scheme_XRR}
\end{figure}

The presence and evolution of the MG layer at the air/water interface suggest that it originates from the MG cores; to be precise from the part of the cores that float on the apparent air/water interface, as suggested earlier by neutron reflectometry and corresponding MD-simulations.\cite{2022Bochenek, 2024Gerelli} 
This view is further corroborated by the AFM images presented in Figure~\ref{pic:DIP_AFM_OSR}C. 
Upon uniaxial compression, the area per MG is reduced steadily, and as a consequence neighbouring MG particles need to interpenetrate each other and/or deform, presumably diving into the water subphase. 
The minimum distance of adjacent MG particles is found \textit{via} \textit{ex situ} AFM measurements to be \SI{416}{nm} at $\Pi\,=\,\SI{30}{mNm^{-1}}$ (Figure~S1.1 in the SI). We highlight that this value is below the MG hydrodynamic diameter of \SI{616}{nm} in bulk at \SI{20}{\celsius}, which is still considerable larger than its value in the collapsed state of \SI{269}{nm} at \SI{50}{\celsius} as measured with DLS.
Provided that the MG particles remain in a monolayer, they must deform when the pressure increases. However, to which extent the water fraction in the monolayer varies remains unresolved using XRR alone, due to the lack of contrast between the MG particles and the H$_2$O subphase. 
Nevertheless, compressing the MG particles from $\Pi\,=\,\SI{17}{mNm^{-1}}$ to $\Pi\,=\,\SI{22}{mNm^{-1}}$ (i.e., decreasing $D_\mathrm{cc}$) yields an almost constant $h_\mathrm{MG}$, which indicates a release of water from the MG particles. A de-swelling upon compression is also in line with the observed reduction in the length and roughness of the air-adjacent layer. 
In this case, we hypothesise that the MG particles will (i) adopt their initial ``fried egg'' shape in a hexagonal pattern in the sample plane (faceting), (ii) release water molecules and (iii) migrate deformable parts perpendicular to the air/water interface upon compression.
Indeed, the latter becomes particularly visible for points in the isotherm of highest curvature: $\Pi\,\leq\,\SI{10}{mNm^{-1}}$ and $\Pi\,\ge\,\SI{24}{mNm^{-1}}$.  
We suggest that the reason can be twofold: Upon compression,
\begin{enumerate}[a)]
    \item the water is squeezed out from the MG particles forcing a 3D shrinkage, both vertically and horizontally, and
    \item the dangling ends of the MG particles increase the thickness of the homogenous layer between the cores of the MG particles. This results in a \textit{pseudo} decrease in the distance between the apparent air/water interface and pole-cap of the floating MG (see Figure~\ref{pic:Scheme_XRR}).
\end{enumerate}
A complementary study by \citeauthor{2018Wu} provides insights into hypothesis (a), demonstrating the submersion of hard silica spheres upon lateral compressing \textit{via} XRR and grazing incidence small angle X-ray scattering (GISAXS).\cite{2018Wu}
Here the authors explained submersion of the silica particles by the asymmetric electrostatic repulsion between equally charged neighbouring particles at the air/water interface due to the high dielectric constant of water.\cite{2018Wu} 
However, with respect to the second hypothesised reason (b), the meniscus (water distribution) around the floating MG particles and each other approaching MG particle must be considered as well. Unfortunately, however, this is experimentally not possible with X-ray scattering due to the lack of contrast between the MGs and water. 
Here in this work, the amount of scattering material (either water or MG) floating above the air/water interface reduces nearly linearly for $\Pi > $ \SI{10}{mNm^{-1}} (inset of Figure~\ref{pic:XRR}B).
The smooth transition from the low to high SLD region marks the apparent air/water interface and is defined by the point of inflection of the SLD where $z\,=\,\SI{0}{\angstrom}$. 
Upon compression, this region also thins, but the SLD contrast towards bulk water is marginal. 
At the highest applicable lateral pressure, this layer persists, exhibiting an SLD slightly higher than that of bulk water, which may be an artefact arising from non-negligible diffuse scattering.

\subsection{Comparison across \textit{\textit{in situ}} and \textit{ex situ} approaches}

\begin{figure}[h]
\centering
\includegraphics[width=\linewidth]{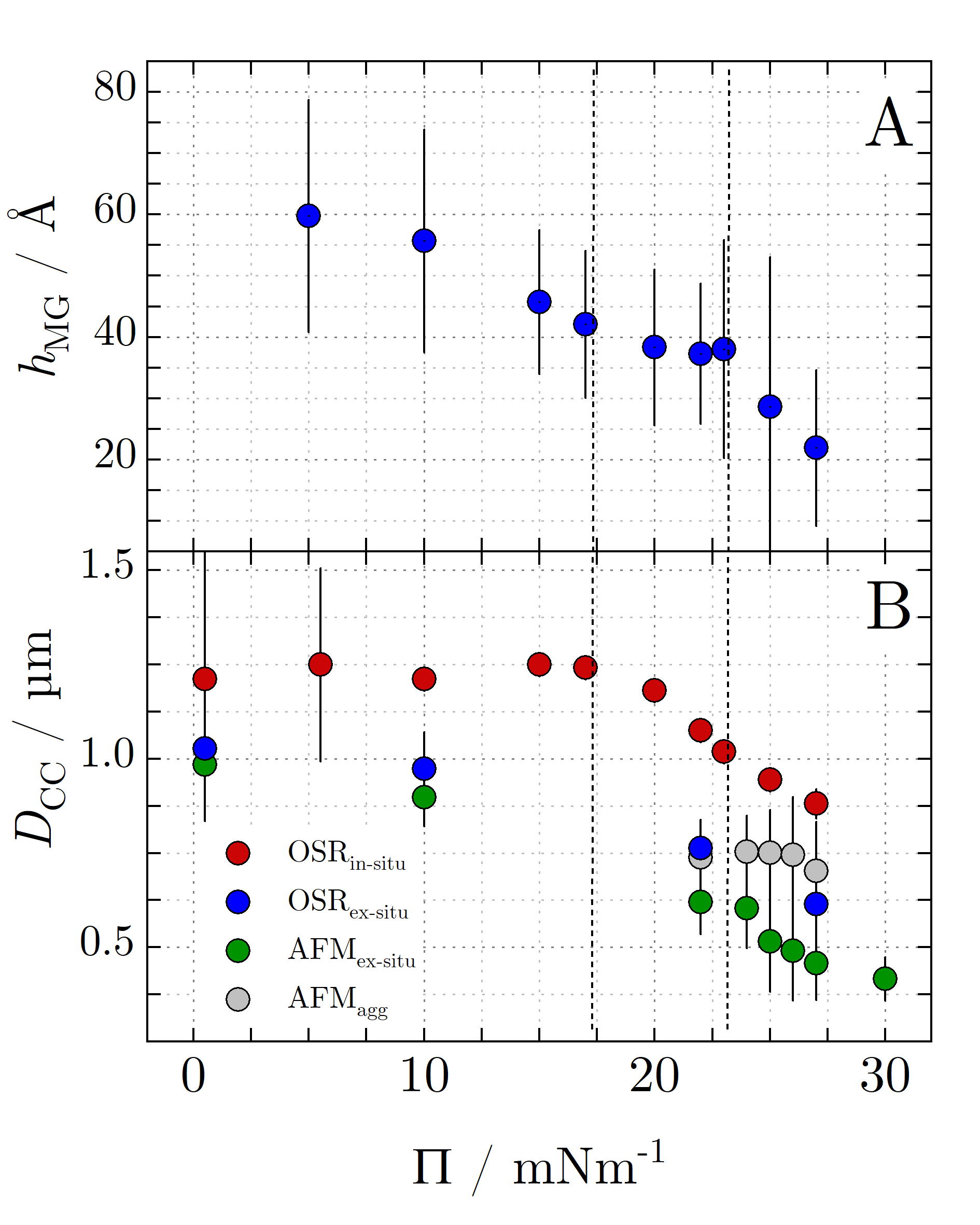}  
\caption{(A) The first moment ($h_\mathrm{MG}$) of the slabs $j = 1, 2$ that represent the parts of the MG that protrude into the air superphase above the apparent air/water interface ($z\,\geq\,0\,\mathrm{\Angstrom}$), as calculated from the SLD profiles presented in Figure~\ref{pic:XRR}. (B) Centre-to-centre distances ($D_\mathrm{cc}$) of MG particles deduced with OSR at the air/water interface (\textit{in situ}) and with AFM and OSR on Langmuir-Blodgett transferred samples (\textit{ex situ}) as a function of lateral pressure $\Pi$.} 
\label{pic:Dcc}
\end{figure}

A comparison across the \textit{ex situ} methods reveals a small systematic deviation in the extracted centre-to-centre distances ($D_\mathrm{cc}$) between the MG particles. 
The OSR values are always larger than those deduced by AFM (see Figure~\ref{pic:Dcc}B), however still within the uncertainty.
This is presumably due to the umbrella effect which is not taken into account here. 
Nevertheless, the comparison clearly shows that OSR is well-suited to determine lattice constants of immobilised MG at the air/solid interface, making it a suitable technique to investigate MG also at other interfaces, such as the gas/liquid interface. 

At the air/water interface, we found that $D_\mathrm{cc}$ are ($\approx$~\SIrange{20}{30}{\percent}) greater than the \textit{ex situ} values when compressed from the lowest to the highest pressure.
However, \textit{in situ} OSR reveals a slightly different picture. 
Below $\Pi\,=\,20\,\mathrm{mNm^{-1}}$, the lattice constant is in a \textit{quasi}-constant state. 
Interestingly, for low to intermediate pressures (\SIrange{0.5}{17}{mNm^{-1}}) a distinct OSR signal with the same peak position is detected, meaning that MG domains are already present and that the lattice constant inside the domains remains constant upon compression. 
This can be interpreted as a coexistence of areas decorated with MG  and a MG free air/water interface. 
In this pressure region, the area which is occupied by MG domains increases with increasing pressure, until the MG particles completely occupy the interface.
Interestingly, however, upon compression the MG particles neither deform nor change their distance from each other. 
Here inter-domain distances decrease whilst intra-domain distances remain constant.
The change in the surface pressure (i.e., interfacial tension) is therefore governed by the area fraction of the domains. 
This view also provides an explanation as to why the surface tension from MG solutions of sufficiently high concentration reduces to values that correspond to the surface pressure of the plateau region of the corresponding Langmuir isotherm.
As a result, the established view on MG structuring in the low-pressure regime ($\Pi\,\leq\,17\,\mathrm{mNm^{-1}}$),  for example that of \citeauthor{Pinaud2014}\cite{Pinaud2014}, needs to be reconsidered. 
Estimating the distance between MG particles across a given area assumes a homogeneous distribution of those MG particles. This is obviously not the case at the air/water interface at lower pressures and results in an averaged value. 
This view is further supported by Figure~S1.2 in the SI, where close packed MG domains coexist with uncovered areas at low pressures. Recently the heterogenous coverage was also observed for gold particle loaded MG that forms domains and do not spread homogenously, even when a larger area is available \cite{Zhou2025}
In that regard, special care should be taken when considering the definition of interfacial elasticity, to be precise the change of area per molecule upon compression.

Upon further compression towards the flattening of the isotherm (Figure~\ref{pic:OSR}B), the lattice constant of the MG particles reduces with decreasing area (Figure~\ref{pic:Dcc}B). 
The plateau can therefore be regarded as a phase of solidification, whereby the MG particles must deform laterally. Interestingly, the XRR results show that the vertical deformation is less pronounced in this pressure region (as indicated in Figure~\ref{pic:Dcc}A), as $h_\mathrm{MG}$ (the first moment of the polymer density) decreases with increasing pressure. 
Above $\Pi\,=\,\SI{23}{mNm^{-1}}$, when the slope of the isotherm increases again, the vertical deformation is more pronounced and the MG particles appear to be either pushed underwater or arranged so close to each other that the MG-caps appear thinner. 
At the highest applicable pressure of $\Pi\,=\,27\,\mathrm{mNm^{-1}}$, the XRR results show a thin, relatively smooth layer which could only be achieved if the MG particles deform from a round, smooth object into a more faceted shape with sharper edges. 
Unfortunately, due to the lateral inhomogeneity of the sample, further interpretation of the XRR results is limited, however, a decreasing layer thickness and changing surface roughness can be reliably extracted. 

In general, the \textit{in situ} $D_\mathrm{cc}$ is larger compared to those deduced by \textit{ex situ} approaches.
This indicates that the immobilisation process \textit{via} LB-transfer on a hydrophilic substrate is invasive and does not preserve the long-range structure of the MGs. 
Specifically, when immobilised, capillary forces induce an attractive force between the MG particles that drives agglomeration on a native oxide silicon wafer. 
This scenario is possible when the MG particles are sufficently soft and mobile, which is the case here for the employed intermediate cross-linked MGs ($\SI{5}{mol\%}$) on a hydrophilic interface.\cite{2024Kuk}
While the $D_\mathrm{cc}$ is constant for $\Pi\,\leq\,\SI{17}{mNm^{-1}}$ at the air/water interface, a slight decrease in $D_\mathrm{cc}$ is observed at the air/solid interface for $\Pi\,\leq\,\SI{10}{\mathrm{mNm^{-1}}}$; this confirms the invasive nature of the LB-transfer in the present case.

Nevertheless, at the highest pressure of $\Pi\,=\,\SI{27}{mNm^{-1}}$, the employed model for the \textit{in situ} OSR data does not describe the OSR signal as well as for the lower pressures. 
The reasons for this can be myriad. However, for this case we propose that (i) the deformation of the MG particles requires a sophisticated treatment of the form factor; (ii) the strong compression leads to the formation of MG bilayers; or (iii) the shells of neighbouring MG particles begin to interpenetrate one another. 
The OSR data presents a drastic phase change upon compression to $\Pi\,=\,\SI{27}{mNm^{-1}}$ that is supported by the isotherm.
However, we unfortunately cannot conclude if the hexagonal phase of the MG particles is maintained or broken with the simple model applied here; \textit{ex situ} AFM results do suggest that the hexagonal phase is maintained.
Further investigation employing techniques with greater contrast between the MG particles and the bulk phase (i.e., neutron scattering techniques) would be required to unravel this phenomenon.

\section{Conclusion}

For the first time, \textit{in situ} off-specular X-ray reflectometry (OSR) is systematically applied to studying the lateral ordering of self -assembled soft matter at the air/water and air-solid interfaces.
OSR revealed hexagonal packing of polymer microgel particles in domains of random orientation. 
The resultant centre-to-centre distances ($D_\mathrm{cc}$) as determined \textit{in situ} by OSR were found to be larger than the respective \textit{ex situ} LB-transferred MGs on (hydrophilic) silicon substrates examined by OSR and atomic force microscopy (AFM). 
These results confirm the earlier findings of Kuk~et~al.\cite{2023Kuk, 2024Kuk} for larger microgels with a silicon core, studied with light scattering and AFM. 
This effect is explained by a weak adhesion of the MG particles towards the hydrophilic substrate, whereby the lateral ordering of MG particles is driven by the attractive capillary forces induced by thin film wetting in the film formation during LB-transfer. 
In contrast to the LB-transferred MG particles, $D_\mathrm{cc}$ is noted to be constant for \textit{in situ} OSR for $\Pi\,\leq\,\SI{17}{mNm^{-1}}$, and reduces only when the pressure exceeds \SI{17}{mNm^{-1}}; across the entire isotherm, the surface pressure monotonically increases with increasing compression. 
That is, in the low $\Pi$ region, inter-domain $D_\mathrm{cc}$ decreases whilst maintaining a constant intra-domain $D_\mathrm{cc}$.
Above the critical pressure of \SI{17}{mN\per m}, the intra-domain $D_\mathrm{cc}$ begins to decrease with increasing $\Pi$.
This strongly suggests that crystalline domains of hexagonally-packed MGs are present at low surface pressures and are pushed together during compression. 
At the highest pressure of $\Pi\,=\,\SI{27}{mNm^{-1}}$, the employed model is no longer sufficient to describe the OSR data, presumably as the MGs are forming a flat layer towards air as indicated by the specular reflectivity. 
This reveals a thinning of the MG-cap upon compression, which points towards either an immersion of the remaining part of the MGs into the adjacent aqueous phase or an increased hydration of the MG particles upon compression. 
However, these phenomena cannot be deconvoluted due to the to lack of contrast between the MGs and the subphase.

By comparing the \textit{ex situ} OSR and AFM results on identical samples, we validate that the scattering technique is quantitatively able to investigate MGs at the air/solid interface. 
We also demonstrate for the first time that \textit{in situ} XRR and OSR are sensitive methods capable of resolving MGs at the air/liquid interface. Both scattering methods (i.e., XRR and OSR) require complex modelling within the distorted wave Born approximation (DWBA) due to the lateral inhomogeneity of the MG film and multiple scattering in OSR. 
In contrast to other methods, such as small angle light scattering (SALS) or fluorescence microscopy, the employed lab-based X-ray techniques require no labelling, are non-destructive and are sensitive to particles smaller than $1\,\mathrm{\mu m}$.


\section{Conflicts of interest}

The authors declare no conflicts of interest. 

\section{Supporting Information}

Supporting information: all X-ray scattering data (OSR and XRR); all analysis tools employed for X-ray measurements; all AFM data; all software required to reproduce the nearest neighbour analysis. See DOI: \href{https://www.doi.org/10.48328/tudatalib-1656}{10.48328/tudatalib-1656}.

\section{Acknowledgement}

HR gratefully acknowledges a Humboldt Research Fellowship from the Alexander von Humboldt Stiftung as well as a Career Bridging Grant of the Technische Universit\"at Darmstadt. CL is thankful for the funding by the DFG (German Research Council) within the project 405629430 "Skalen\"ubergreifende Charakterisierung der Wechselwirkung von Cellulosegrenzfl\"achen mit Polymeren". OS would like to thank Ken Garden for providing the Matlab community with the polarplot3d-function (\url{https://www.mathworks.com/matlabcentral/fileexchange/13200-3d-polar-plot}).


\setlength{\bibsep}{0pt plus 0.3ex}
\small
\bibliography{ms.bib} 
\bibliographystyle{rsc} 
\end{document}


\maketitle


\tableofcontents

\newpage
\section{Additional AFM results}

\begin{figure}[htb!]
    \centering
    \includegraphics[width=0.97\textwidth]{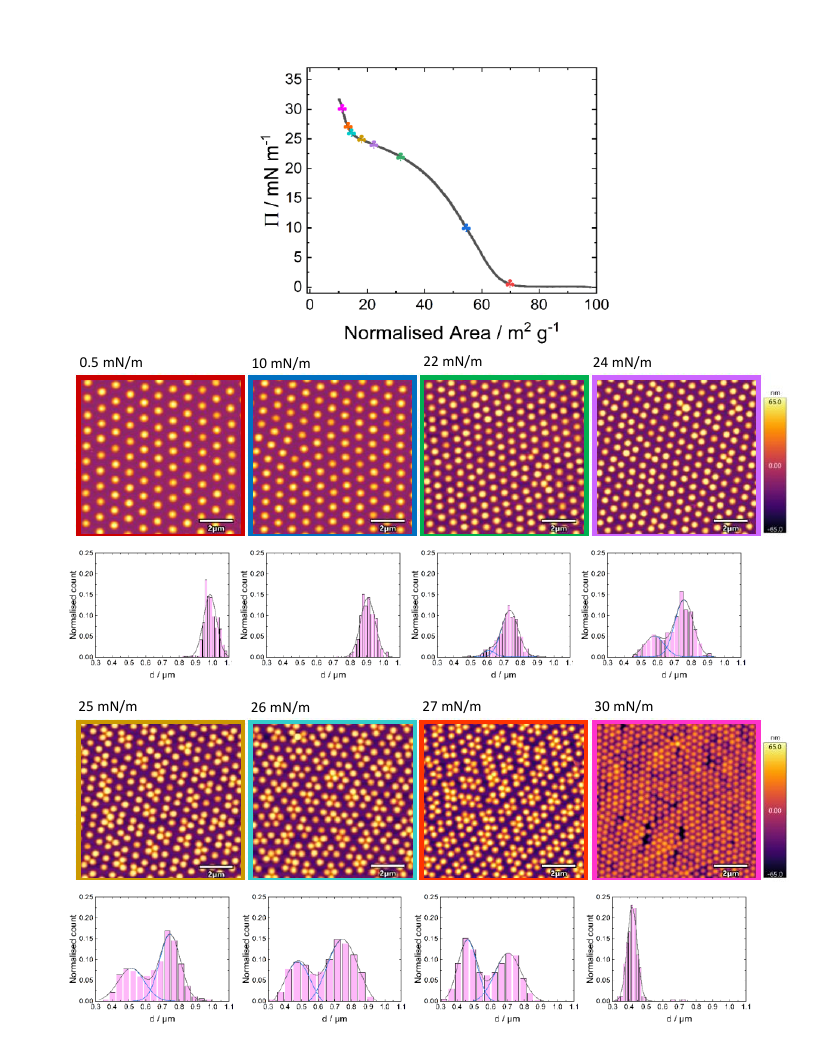}
    \caption{Langmuir compression isotherm of the investigated microgel. The surface pressures $\Pi$ = 0.5, 10, 22, 24, 25, 26, 27 and 30 $\mathrm{mNm^{-1}}$ at which the microgel Langmuir-Blodgett deposition was carried out are marked. The colour code relates to the AFM images at defined surface pressures shown. Below each AFM image, the histograms of the nearest neighbour distance with respective Gauss fits are displayed. Apart from the sample at $\Pi$ = 30$\mathrm{mNm^{-1}}$, where scan sizes of $5\times$\SI{5}{\micro\metre} were used, the evaluation of all other samples was based on scan sizes of $10\times$\SI{10}{\micro\metre}.}
    \label{fig:allAFM}
\end{figure}

Here we present in Figure~\ref{fig:allAFM} the AFM topologies of the LB-transferred MGs for all investigated lateral pressures. 

\begin{figure}[htb!]
    \centering
    \includegraphics[width=\textwidth]{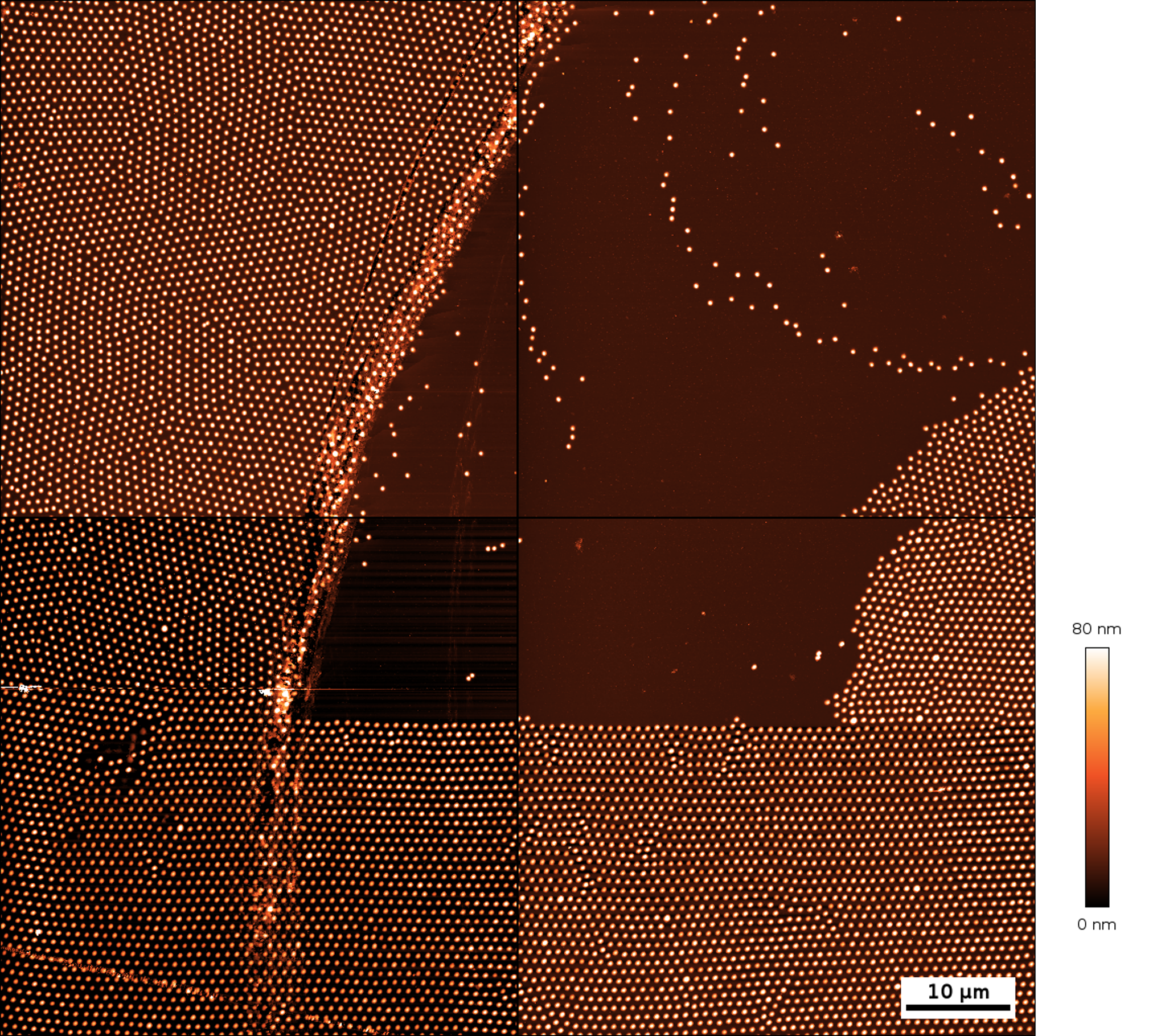}
    \caption{Stitched image from 4 individual AFM scans illustrating the topology of an LB transferred MG at $\Pi\,=\,0.5\,\mathrm{mNm^{-1}}$ over an area of $100\,\times\,100\,\mathrm{\mu m^2}$. Despite the imperfect stitching the images reveals (i) varying orientation of the hexagonal arranged MG over larger distances and (ii) heterogenous distributed MG in domains of closed packed MG and blank areas.}
    \label{fig:stichedAFM}
\end{figure}

\newpage
\section{Additional OSR details}
\setcounter{figure}{0}

As all orientations of the expected 2D-hexagonal lattice are allowed equally in the scattering plane, all possible Bragg peaks have to be taken into account, similar to a powder crystalline sample in an XRD experiment. 
In fact for the lateral structure, the applied geometry is only sensitive in the $q_\mathrm{x}$ direction, but this selection rule is softened since the incoming beam direction is not well defined in the scattering plane. Thus several orientations of the lattice are probed (and integrated) at the same time. This divergence is estimated to be $\Delta 2\theta_{y} = \frac{S_\mathrm{s}\,+\,S_\mathrm{d}}{2L_\mathrm{s,d}}\,\approx\,\SI{3.6}{\degree}$ where $S_\mathrm{s}\,=\,S_\mathrm{d}\,=\,\SI{16}{mm}$, are the openings of the X-ray source and the detector in the $y$-direction and $L_\mathrm{s,d}\,= \,\SI{500}{mm}$ their distance. Taking into account the expected 6 fold symmetry, the probability to hit a ``crystallite'' with the needed orientation reeds $\frac{3.6^\circ\,\cdot\,6}{360^\circ}\,\approx\,6\%$. Therefore, on average every 17th crystallite contributes to the scattering.

Moreover, the orientation of MG particles at the air/water interface is not constant in time and cannot be controlled. This situation is inherently different for LB-transferred MG particles which can be easily controlled \textit{via} rotation during an OSR scan. We were able to utilise this and generate orientation-averaged OSR scans that we compared 1:1 to the AFM scans. 

\subsection{Fitting approach: 3-4 Gaussians with a hemi-ellipsoid form factor}

\begin{figure}
    \centering
    \includegraphics[width=\textwidth]{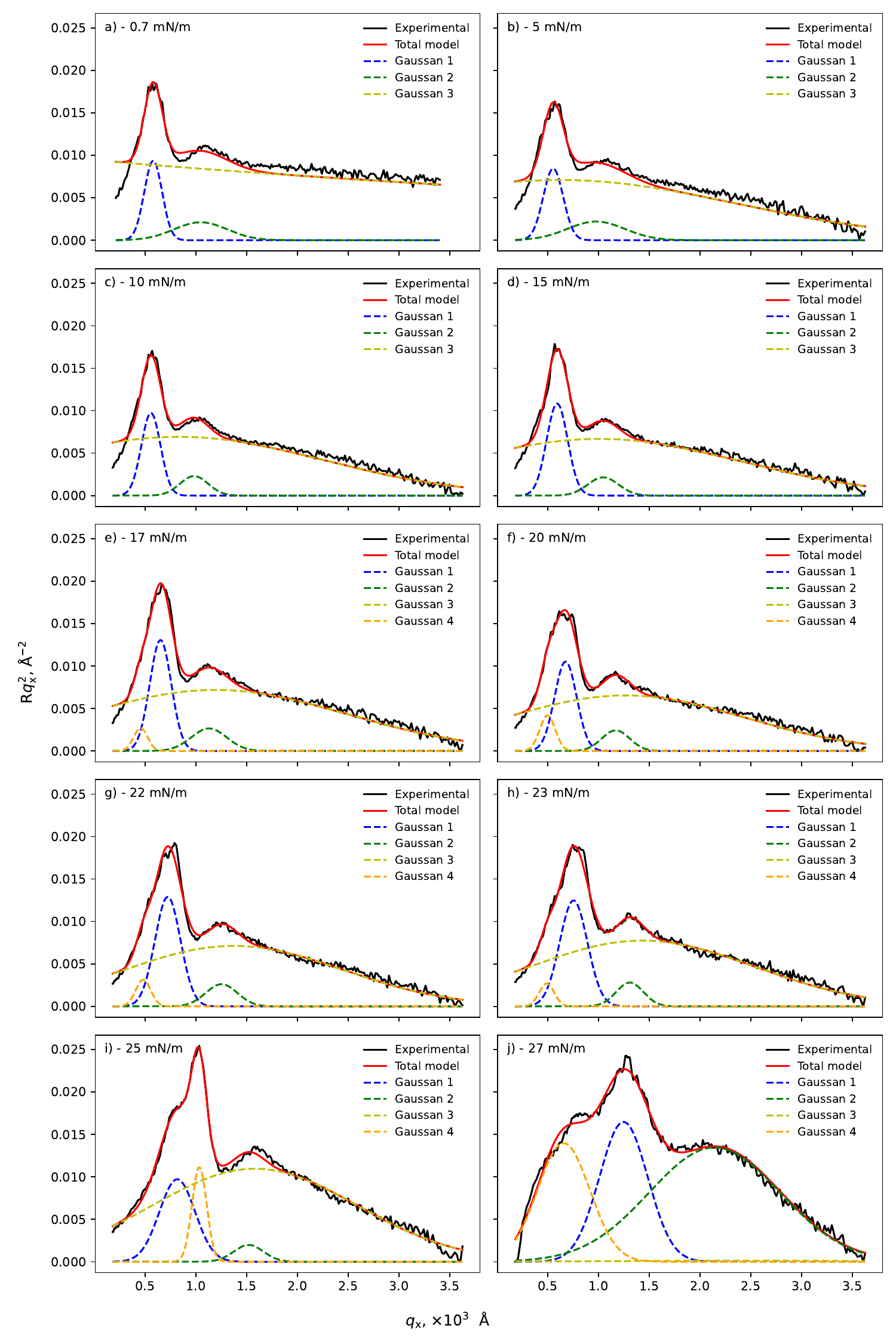}
    \caption{Off-specular reflectivities at various surface pressures with fit that individual components of the overall curve.}
    \label{osr_fits}
\end{figure}

As noted in \S 4.2, our attempt to model the \textit{in situ} OSR employed either 3 or 4 Gaussian functions as a structure factor. Despite the lack of scattering information in the $q_\mathrm{x}$ and $q_\mathrm{z}$ plane we here tested a form factor for the MG-particles that is a hemi-ellipsoid using the BornAgain software package.  
The results are presented in Figure~\ref{osr_fits}.

The centre positions of each Gaussian function within the structure factor were constrained (2\% variation) to adhere to the 2D hexagonal lattice spacing, with position ratios of $1, \sqrt{3}, 2$.
From a surface pressure of \SI{15}{mN/m} onwards, a fourth Gaussian was required to adequately model the data; this was not constrained but included to capture the shoulder at low $q$.

\newpage

\newpage
\begin{landscape}
\subsubsection{OSR parameters}
\renewcommand{\arraystretch}{1.2}
\pagestyle{empty}
\begin{table}[htb!]
\caption{Parameters relating to the OSR model which consists of a series of 3 or 4 Gaussian functions where the peak positions are related by $1, \sqrt{3}, 2$.}
\footnotesize
\begin{tabular}{l|ll|ll|ll|ll}
Surface pressure, & \multicolumn{2}{l|}{Gaussian 1} & \multicolumn{2}{l|}{Gaussian 2} & \multicolumn{2}{l|}{Gaussian 3} & \multicolumn{2}{l}{Gaussian 4} \\
mN/m              & Center, \si{\angstrom}          & Sigma, \si{\angstrom}         & Center, \si{\angstrom}          & Sigma, \si{\angstrom}         & Center, \si{\angstrom}          & Sigma, \si{\angstrom}         & Center, \si{\angstrom}         & Sigma, \si{\angstrom}         \\ \hline
0.7 & 5.812 $\pm$ 0.033 & 0.882 $\pm$ 0.006 & 10.543 $\pm$ 0.072 & 2.614 $\pm$ 0.018 & 12.177 $\pm$ 0.075 & 42767 $\pm$ 287.83 & - & -  \\
5 & 5.569 $\pm$ 0.034 & 0.985 $\pm$ 0.007 & 9.899 $\pm$ 0.064 & 2.825 $\pm$ 0.020 & 11.428 $\pm$ 0.076 & 17.457 $\pm$ 0.118 & - & -  \\
10 & 5.598 $\pm$ 0.038 & 0.946 $\pm$ 0.006 & 9.859 $\pm$ 0.069 & 1.393 $\pm$ 0.009 & 11.402 $\pm$ 0.074 & 14.202 $\pm$ 0.096 & - & -  \\
15 & 5.954 $\pm$ 0.042 & 1.037 $\pm$ 0.007 & 10.490 $\pm$ 0.072 & 1.510 $\pm$ 0.010 & 12.106 $\pm$ 0.074 & 13.901 $\pm$ 0.090 & - & -  \\
17 & 6.533 $\pm$ 0.041 & 1.063 $\pm$ 0.007 & 11.313 $\pm$ 0.073 & 1.743 $\pm$ 0.011 & 13.072 $\pm$ 0.088 & 12.900 $\pm$ 0.086 & 4.572 $\pm$ 0.029 & 0.682 $\pm$ 0.005  \\
20 & 6.782 $\pm$ 0.046 & 1.087 $\pm$ 0.008 & 11.719 $\pm$ 0.083 & 1.369 $\pm$ 0.008 & 13.568 $\pm$ 0.089 & 11.653 $\pm$ 0.076 & 4.993 $\pm$ 0.033 & 0.764 $\pm$ 0.005  \\
22 & 7.249 $\pm$ 0.051 & 1.260 $\pm$ 0.008 & 12.550 $\pm$ 0.076 & 1.557 $\pm$ 0.010 & 14.501 $\pm$ 0.095 & 10.722 $\pm$ 0.075 & 4.828 $\pm$ 0.036 & 0.746 $\pm$ 0.005  \\
23 & 7.568 $\pm$ 0.051 & 1.366 $\pm$ 0.010 & 13.102 $\pm$ 0.078 & 1.338 $\pm$ 0.009 & 15.154 $\pm$ 0.105 & 11.027 $\pm$ 0.069 & 4.916 $\pm$ 0.034 & 0.739 $\pm$ 0.005  \\
25 & 8.187 $\pm$ 0.057 & 1.754 $\pm$ 0.011 & 15.201 $\pm$ 0.091 & 1.462 $\pm$ 0.010 & 16.237 $\pm$ 0.104 & 10.066 $\pm$ 0.071 & 10.350 $\pm$ 0.054 & 0.723 $\pm$ 0.004  \\
27 & 12.493 $\pm$ 0.087 & 2.376 $\pm$ 0.013 & 21.581 $\pm$ 0.132 & 6.444 $\pm$ 0.044 & 24.962 $\pm$ 0.178 & 11.780 $\pm$ 0.081 & 6.604 $\pm$ 0.039 & 2.572 $\pm$ 0.019  \\
\end{tabular}
\end{table}
\vfill\raisebox{-4em}{\makebox[\linewidth]{\thepage}}
\end{landscape}    

\subsection{Fitting approach: $2\times3$ Gaussians}

\begin{figure}[htb!]
    \centering
    \includegraphics[width=\textwidth]{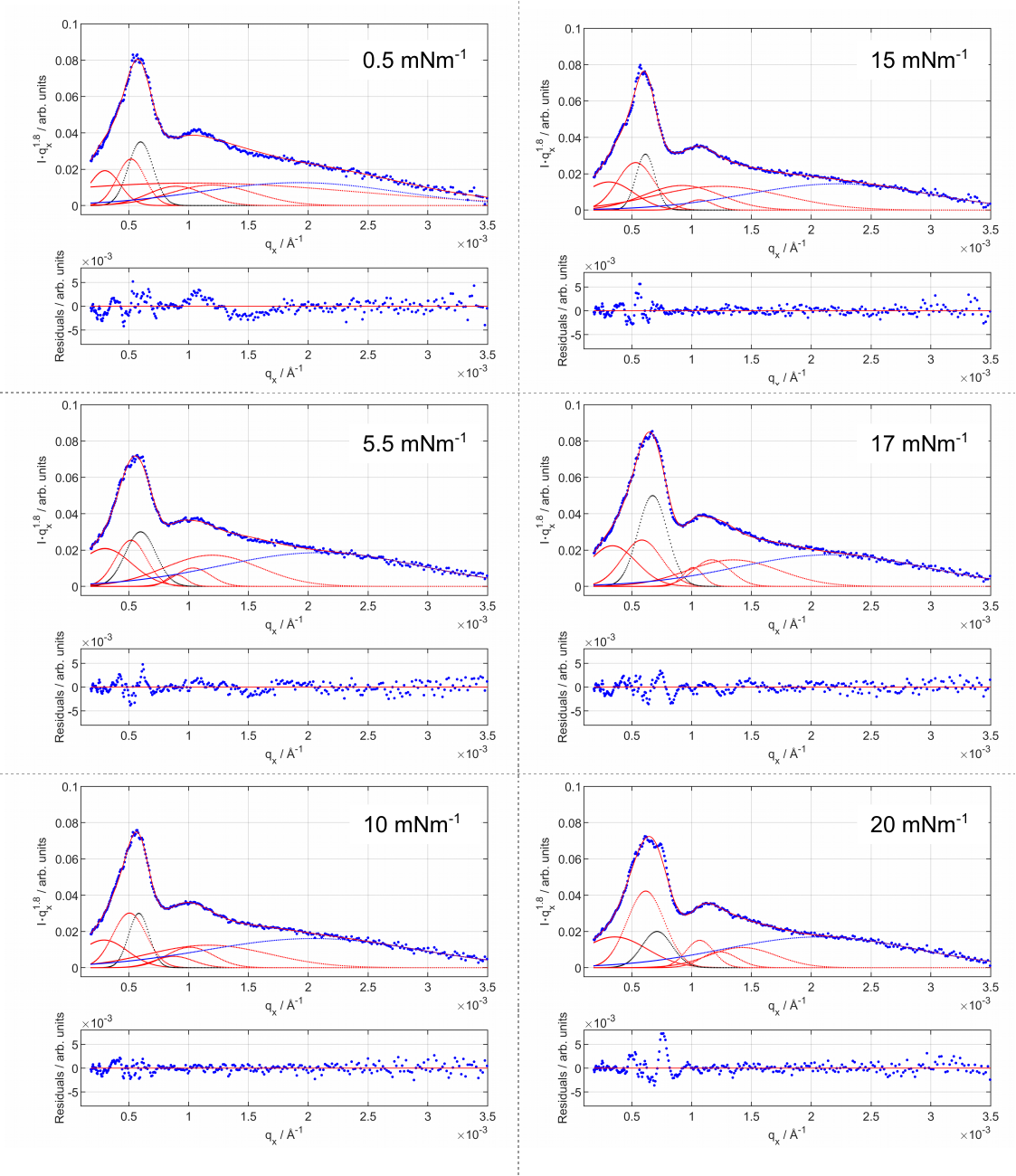}
    \caption{Off-specular reflectivities at surface pressures up to $20\,\mathrm{mNm^{-1}}$ with corresponding fits. The black Gaussian refers to the lowest indexed peak, the umbrella effect is addressed by the red Gaussians whose positions are fixed by the ratios $\sqrt{1/4}$ and $\sqrt{3/4}$ for the $1^\mathrm{st}$ order and $2\sqrt{1/4}$, $2$ and $2\sqrt{3/4}$ for the $2^{nd}$ order. The form factor is represented by the blue curve. Below the reflectivity the residuals are presented}
    \label{Collage_OSR_Insitu_1}
\end{figure}
\newpage

\begin{figure}[htb!]
    \centering
    \includegraphics[width=\textwidth]{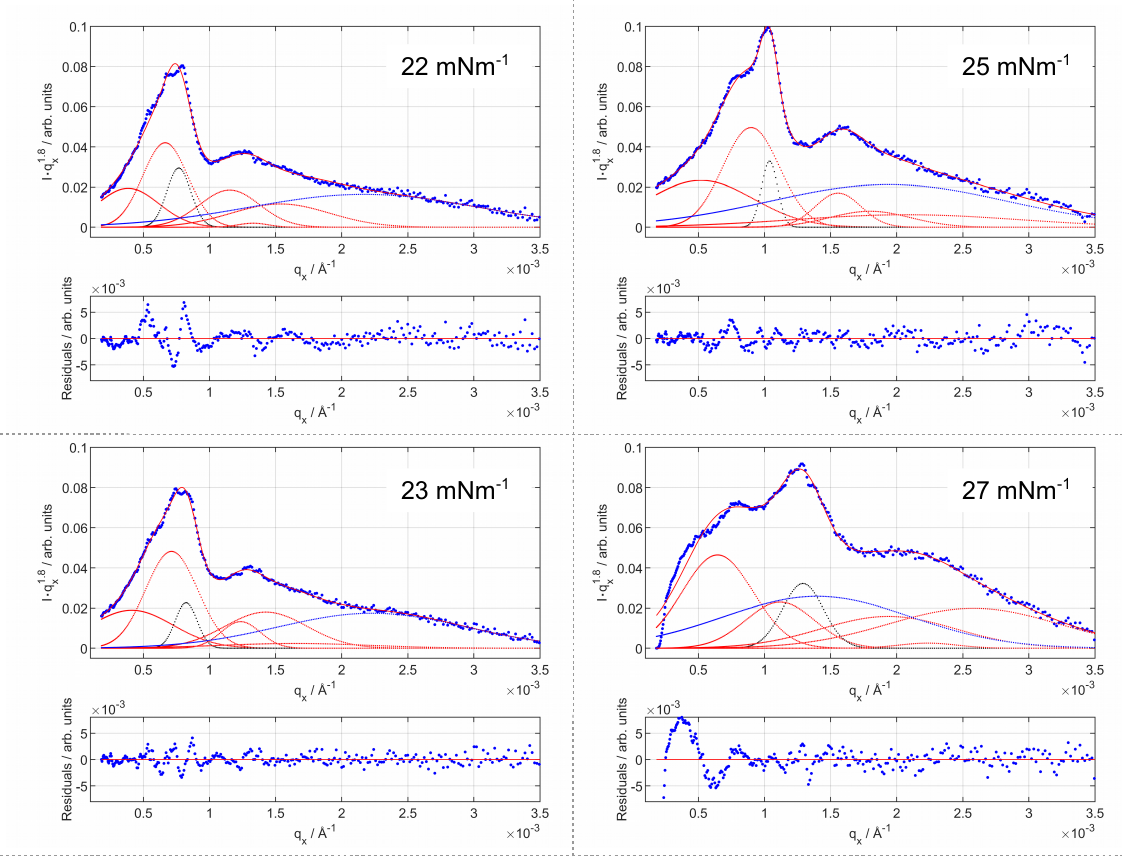}
    \caption{Off-specular reflectivities at surface pressures from above $22\,\mathrm{mNm^{-1}}$ with corresponding fits. The black Gaussian refers to the lowest indexed peak, the umbrella effect is addressed by the red Gaussians whose positions are fixed by the ratios $\sqrt{1/4}$ and $\sqrt{3/4}$ for the $1^\mathrm{st}$ order and $2\sqrt{1/4}$, $2$ and $2\sqrt{3/4}$ for the $2^{nd}$ order. The form factor is represented by the blue curve. Below the reflectivity the residuals are presented}
    \label{Collage_OSR_Insitu_2}
\end{figure}

\newpage

\subsection{Reciprocal space map}

\begin{figure}[htb!]
    \centering
    \includegraphics[width=\textwidth]{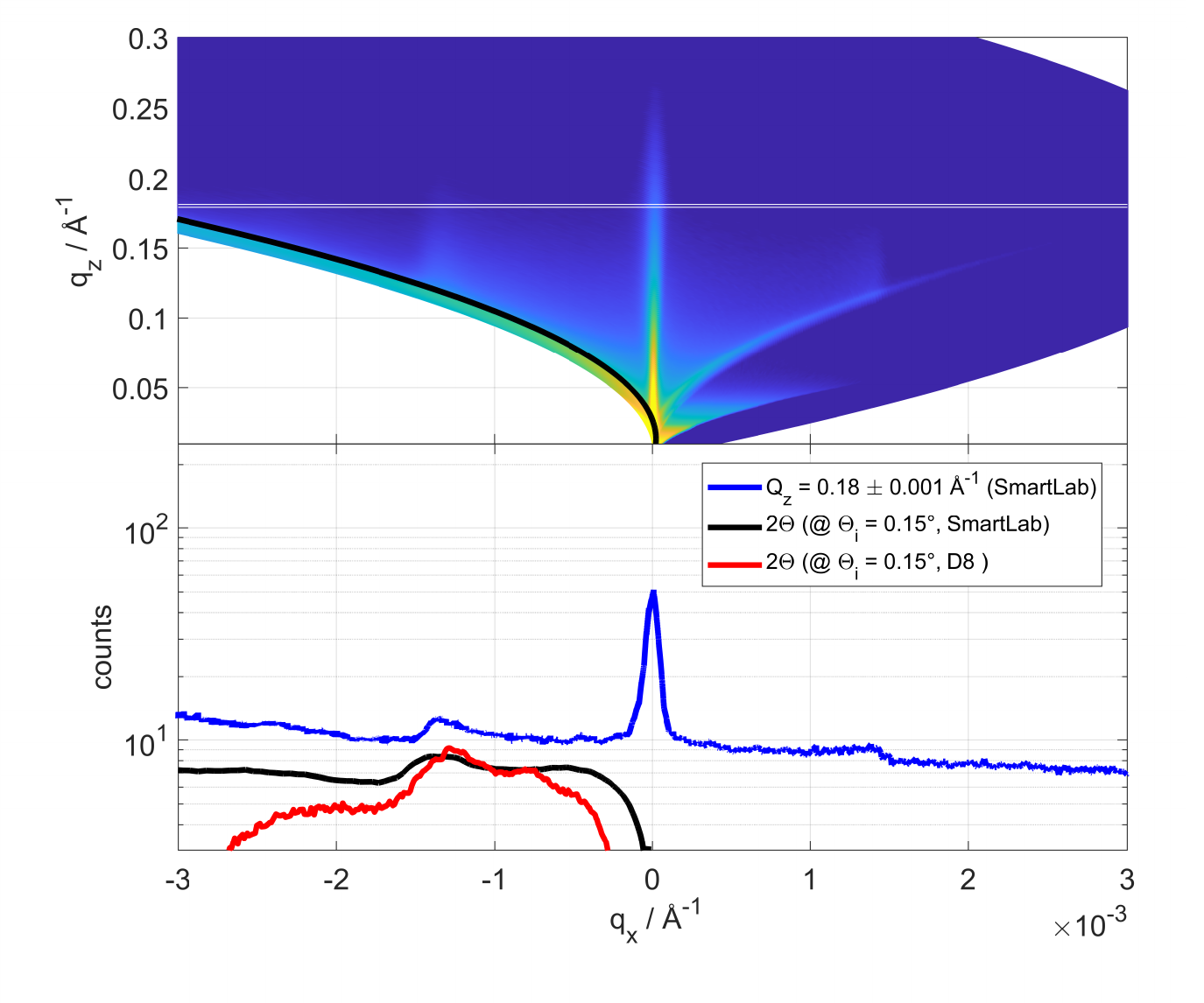}
    \caption{Reciprocal space map of MG particles at the air/water interface at $\Pi\,=\SI{27}{mNm^{-1}}$ (top) and extracted cuts along constant $q_\mathrm{z}$ ($\mathrm{\theta_i} + \mathrm{\theta_f}\,=\,\mathrm{const.}\,\approx\,\SI{2.53}{\degree}$, blue), constant incident angle ($\mathrm{\theta_i}\,=\,\SI{0.15}{\degree}$, black) (bottom). The scattering curve from the manuscript is here plotted in red for comparison. }
    \label{RSM}
\end{figure}

Here we present a reciprocal space map (RSM) of PNIPAM microgel particles with 5~mol\% cross-linker at the air/water interface (Figure~\ref{RSM}).
The reciprocal space mapping was carried out on the Rigaku SmartLab using the same trough as explained in \S3.4 of the manuscript.
In particular, Figure~\ref{RSM} demonstrates the clear symmetry in the off-specular signal, and that the Bragg peaks are not dependent on $q_\mathrm{z}$; i.e., there is no bending in the peaks at $q_\mathrm{x} \approx \pm$\SI{1.4e-3}{\per\angstrom}.
The extracted cut along $q_\mathrm{z} =$~\SI{0.18}{\per\angstrom} (blue, Figure~\ref{RSM} lower) demonstrates good agreement with both the extracted cut from the RSM at a constant incident angle of $\theta_\mathrm{i} = $~\SI{0.15}{\degree} from the SmartLab device (black, identical sample) and from the single incident angle of $\theta_\mathrm{i} = $~\SI{0.15}{\degree} from the D8 device (red, non-identical sample).

\newpage
\begin{landscape}
\pagestyle{empty}
\section{XRR model results}
\setcounter{figure}{0}
\setcounter{table}{0}

\renewcommand{\arraystretch}{1.2}
\begin{table}[htb!]
\caption{Parameters derived from optimised models of XRR profiles. From fronting to backing, the interface was modelled as \texttt{air | layer 1 | layer 2 | layer 3 | layer 4 | bulk} using slabs, where \texttt{bulk} refers to the silicon backing. Roughness corresponds to the interface above the layer.}
\footnotesize
\begin{tabular}{l|lll|lll|lll}
\textbf{Surface pressure}, & \multicolumn{3}{c|}{\textbf{Layer 1}}                                                                                                                    & \multicolumn{3}{c|}{\textbf{Layer 2}}                                                                                                                    & \multicolumn{3}{c}{\textbf{Layer 3}}                                                                                                                    \\
mN/m                       & \multicolumn{1}{c}{Thickness, \si{\angstrom}} & \multicolumn{1}{c}{Roughness, \si{\angstrom}} & \multicolumn{1}{c|}{SLD, \si{\times 10^6\angstrom^{-1}}} & \multicolumn{1}{c}{Thickness, \si{\angstrom}} & \multicolumn{1}{c}{Roughness, \si{\angstrom}} & \multicolumn{1}{c|}{SLD, \si{\times 10^6\angstrom^{-1}}} & \multicolumn{1}{c}{Thickness, \si{\angstrom}} & \multicolumn{1}{c}{Roughness, \si{\angstrom}} & \multicolumn{1}{c}{SLD, \si{\times 10^6\angstrom^{-1}}} \\ \hline
5                          & 82.63 $\pm$ 21.92                             & 41.32 $\pm$ 10.96                            & 0.92 $\pm$ 0.63                                          & 82.63 $\pm$ 21.92                             & 41.32 $\pm$ 10.96                            & 3.19 $\pm$ 0.75                                          & 12.35 $\pm$ 2.62                             & 6.18 $\pm$ 1.31                           & 12.02 $\pm$ 0.75                                         \\
10                         & 78.62 $\pm$ 21.95                             & 39.31 $\pm$ 10.98                            & 0.88 $\pm$ 0.60                                          & 78.62 $\pm$ 21.95                             & 39.31 $\pm$ 10.98                            & 3.32 $\pm$ 0.76                                          & 11.83 $\pm$ 3.79                             & 5.92 $\pm$ 1.90                             & 11.78 $\pm$ 0.79                                         \\
15                         & 65.95 $\pm$ 13.25                             & 32.98 $\pm$ 6.63                            & 0.83 $\pm$ 0.58                                          & 65.95 $\pm$ 13.25                             & 32.98 $\pm$ 6.63                            & 3.47 $\pm$ 0.51                                          & 9.33 $\pm$ 1.40                             & 4.67 $\pm$ 0.70                            & 11.27 $\pm$ 0.45                                         \\
17                         & 60.38 $\pm$ 13.72                             & 30.19 $\pm$ 6.86                              & 0.84 $\pm$ 0.63                                          & 60.38 $\pm$ 13.72                             & 30.19 $\pm$ 6.86                              & 3.41 $\pm$ 0.59                                          & 8.69 $\pm$ 1.92                              & 4.34 $\pm$ 0.96                              & 11.18 $\pm$ 0.53                                         \\
20                         & 55.94 $\pm$ 14.12                             & 27.97 $\pm$ 7.06                              & 0.73 $\pm$ 0.70                                          & 55.94 $\pm$ 14.12                             & 27.97 $\pm$ 7.06                              & 3.21 $\pm$ 0.65                                          & 13.95 $\pm$ 31.69                             & 6.98 $\pm$ 15.85                            & 11.22 $\pm$ 1.91                                         \\
22                         & 53.90 $\pm$ 12.97                              & 26.95 $\pm$ 6.45                             & 0.73 $\pm$ 0.62                                          & 53.9 $\pm$ 12.97                              & 26.95 $\pm$ 6.49                             & 3.09 $\pm$ 0.53                                          & 13.69 $\pm$ 24.39                             & 6.85 $\pm$ 12.20                            & 11.12 $\pm$ 1.40                                         \\
23                         & 54.01 $\pm$ 22.57                             & 27.01 $\pm$ 11.29                           & 0.79 $\pm$ 0.71                                          & 54.01 $\pm$ 22.57                             & 27.01 $\pm$ 11.29                           & 3.09 $\pm$ 0.55                                          & 22.1 $\pm$ 18.18                              & 11.05 $\pm$ 9.09                              & 11.98 $\pm$ 1.94                                         \\
25                         & 35.97 $\pm$ 27.91                             & 17.99 $\pm$ 13.96                           & 1.44 $\pm$ 1.46                                          & 35.97 $\pm$ 27.91                             & 17.99 $\pm$ 13.96                           & 3.45 $\pm$ 3.33                                          & 26.87 $\pm$ 48.70                              & 13.44 $\pm$ 24.35                            & 13.35 $\pm$ 0.82                                        
\end{tabular}
\end{table}

\vspace{2em}
\begin{table}[htb!]
\footnotesize
\begin{tabular}{l|lll|ll}
\textbf{Surface pressure}, & \multicolumn{3}{c|}{\textbf{Layer 4}}                                                                                                                    & \multicolumn{2}{c}{\textbf{Bulk}}    \\
mN/m                       & \multicolumn{1}{c}{Thickness, \si{\angstrom}} & \multicolumn{1}{c}{Roughness, \si{\angstrom}} & \multicolumn{1}{c|}{SLD, \si{\times 10^6\angstrom^{-1}}} & Roughness, \si{\angstrom}         & SLD, \si{\times 10^6\angstrom^{-1}}              \\ \hline
5                          & 460.7 $\pm$ 222.2                             & 33.41 $\pm$ 13.22                             & 14.54 $\pm$ 0.17                                         & 33.41 $\pm$ 13.22 & 14.53 $\pm$ 0.00 \\
10                         & 390.6 $\pm$ 202.8                             & 31.0 $\pm$ 14.34                              & 14.54 $\pm$ 0.16                                         & 31.0 $\pm$ 14.34  & 14.53 $\pm$ 0.00 \\
15                         & 319.1 $\pm$ 49.31                             & 23.37 $\pm$ 5.92                             & 14.42 $\pm$ 0.13                                         & 23.37 $\pm$ 5.92 & 14.53 $\pm$ 0.00 \\
17                         & 284.5 $\pm$ 43.80                              & 22.26 $\pm$ 6.82                             & 14.40 $\pm$ 0.14                                         & 22.26 $\pm$ 6.82 & 14.53 $\pm$ 0.00 \\
20                         & 220.0 $\pm$ 34.09                             & 19.64 $\pm$ 15.0                              & 14.31 $\pm$ 0.15                                         & 19.64 $\pm$ 15.0  & 14.53 $\pm$ 0.00 \\
22                         & 173.3 $\pm$ 19.97                             & 19.14 $\pm$ 8.276                             & 14.14 $\pm$ 0.15                                         & 19.14 $\pm$ 8.276 & 14.53 $\pm$ 0.00 \\
23                         & 137.9 $\pm$ 17.60                              & 15.72 $\pm$ 8.35                             & 14.17 $\pm$ 0.28                                         & 15.72 $\pm$ 8.35 & 14.53 $\pm$ 0.00 \\
25                         & 73.70 $\pm$ 50.35                              & 5.39 $\pm$ 7.63                                & 14.23 $\pm$ 0.88                                         & 5.39 $\pm$ 7.63    & 14.53 $\pm$ 0.88
\end{tabular}
\end{table}
\vfill\raisebox{-4em}{\makebox[\linewidth]{\thepage}}
\end{landscape}

